\documentclass[%
superscriptaddress,
nofootinbib,
nobibnotes,
pra,
]{revtex4-2}

\usepackage{float,graphicx}
\usepackage{dcolumn}
\usepackage{bm}
\usepackage{cancel,soul,ulem}

\usepackage{graphicx}
\usepackage{amssymb,amsmath}
\usepackage[utf8]{inputenc}
\usepackage{multirow}
\usepackage{xcolor}
\usepackage{hyperref}
\hypersetup{colorlinks=true,linkcolor=blue,anchorcolor=blue,citecolor=green,filecolor=blue,urlcolor=magenta,bookmarksnumbered=true}
\usepackage[flushleft]{threeparttable}

\usepackage{amsmath}

\usepackage{appendix}

\usepackage{caption}

\usepackage{mathtools}
\usepackage{dsfont}

\usepackage{amssymb}
\usepackage{stackrel}
\usepackage{graphicx}
\usepackage{subcaption}

\usepackage{braket}

\begin{document}


\title{Nuclear scattering via quantum computing} 

\author{Peiyan Wang}
\affiliation{Institute of Modern Physics, Chinese Academy of Sciences, Lanzhou 730000, China}
\affiliation{School of Nuclear Science and Technology, University of Chinese Academy of Sciences, Beijing 100049, China}

\author{Weijie Du}
\email[Email:]{\ duweigy@gmail.com}
\affiliation{Department of Physics and Astronomy, Iowa State University, Ames, Iowa 50010, USA}

\author{Wei Zuo}
\affiliation{Institute of Modern Physics, Chinese Academy of Sciences, Lanzhou 730000, China}
\affiliation{School of Nuclear Science and Technology, University of Chinese Academy of Sciences, Beijing 100049, China}


\author{James P. Vary}
\affiliation{Department of Physics and Astronomy, Iowa State University, Ames, Iowa 50010, USA}

\date{\today}

\begin{abstract}
	We propose a hybrid quantum-classical framework to solve the elastic scattering phase shift of two well-bound nuclei in an uncoupled channel.
	Within this framework, we develop a many-body formalism in which the continuum scattering states of the two colliding nuclei are regulated by a weak external harmonic oscillator potential with varying strength. 
	Based on our formalism, we propose an approach to compute the eigenenergies of the low-lying scattering states of the relative motion of the colliding nuclei as a function of the oscillator strength of the confining potential.
	Utilizing the modified effective range expansion, we extrapolate the elastic scattering phase shift of the colliding nuclei from these eigenenergies to the limit when the external potential vanishes.
	In our hybrid approach, we leverage the advantage of quantum computing to solve for these eigenenergies from a set of many-nucleon Hamiltonian eigenvalue problems.
	These eigenenergies are inputs to classical computers to obtain the phase shift.
	We demonstrate our framework with two simple problems, where we implement the rodeo algorithm to solve the relevant eigenenergies with the IBM Qiskit quantum simulator. 
	The results of both the spectra and the elastic scattering phase shifts agree well with other theoretical results. 
	
\end{abstract}

\maketitle

\section{Introduction}

The scattering between many-nucleon systems plays important roles in our understanding of fundamental symmetries of nature \cite{Epelbaum:2019zqc, Krebs:2020pii} as well as in our knowledge of stellar formation and evolution \cite{Thompson2009, PhysRevLett.82.4176, Nunes:2020bue, RevModPhys.83.195, Johnson:2019sps}. 
However, first principles calculations of nuclear scattering are in general computationally challenging \cite{PhysRevLett.99.022502,PhysRevLett.101.092501,PhysRevC.79.014610,Elhatisari:2015iga}
due to the strong, non-perturbative interactions \cite{Preskill2018quantumcomputingin}. 

Quantum computing holds the promise to address the demands for computational resources needed for first-principles calculations of quantum many-body problems \cite{Feynman1982,lee2023evaluating}. It leverages the principles of quantum mechanics and holds the promise to address the difficulties encountered by the classical computing techniques by utilizing the coherent superposition and entanglement intrinsic to quantum hardware \cite{nielsen2010quantum}. The utility of quantum computing has been explored in fields such as the quantum chemistry \cite{cao2019quantum,McArdle_2020,bauer2020quantum,Beck:2023xhh}, quantum field theory \cite{Jordan_2012,PhysRevA.98.032331, jordan2019quantum, PhysRevD.101.074512,PhysRevD.106.054508,bauer2022quantum,PhysRevLett.131.081901,PRXQuantum.4.030323}, condensed matter physics \cite{PhysRevA.92.062318,PRXQuantum.2.030307,PhysRevResearch.4.023097} and many other areas \cite{dalzell2023quantum}.
The advent of the era of quantum advantage is also celebrated in the field of low-energy nuclear physics.
Various algorithms \cite{Du:2021ctr,P_rez_Obiol_2023,lv2022qcsh,PhysRevC.105.064308,PhysRevC.105.064318,yang2023shadowbased} have been proposed and implemented to solve the structure properties of nuclear systems via quantum computers. Prototype investigations have also been performed on real-world quantum hardware for low-energy nuclear structure theory \cite{PhysRevLett.120.210501,PhysRevC.106.034325,PhysRevC.105.064317}. However, fewer quantum algorithms \cite{Du:2020glq,Turro:2023dhg,Du:2023bpw} have been proposed to study the scattering properties of many-nucleon systems. 
To date, the preparation of the scattering states for general scattering systems that are of specific functional behaviors is still an open question in quantum computing. It appears to be nontrivial to obtain scattering observables based on such scattering states on quantum computers.

In this work, we propose a hybrid quantum-classical framework to solve the phase shift of the elastic scattering between two well-bound nuclei in an uncoupled channel. According to the kinematics, the stable nuclei, which can be neutron ($n$), proton ($p$), and the close-shell nucleus such as $^4$He ($\alpha$), and $^{16}$O, remain at their ground states following elastic scattering.
To the best of our knowledge, very few works \cite{sharma2023scattering} have explored solving for the elastic scattering phase shift via quantum computing.
The major difficulty in solving the phase shift via quantum computing is that the solution of phase requires the information of the probability amplitudes of the scattering wave function, while the projective measurement applied in quantum computing provides only the probability, leaving the phase of the wave function unspecified. 
We take an alternative approach and base our method on the successful techniques of quantum eigensolvers \cite{Kandala_2017,O_Brien_2019,Motta_2019,PhysRevLett.127.040505,qian2021demonstration, beelindgren2022rodeo,Tilly_2022}, and the formula of the modified effective-range expansion (MERE) \cite{busch1998two, Stetcu_2007_PRA, Suzuki_2009, STETCU20101644, PhysRevC.82.034003,Blume_2012,PhysRevC.101.051602} that describes the full energy dependence of the scattering phase shift when the system is confined in an external harmonic oscillator (HO) potential.

Within our framework, we develop a formalism to regulate the inter-nucleus interaction of the colliding nuclei by a weak external harmonic oscillator potential with varying strength, with which the scattering states of the system are discretized.
We then suggest a method to compute the eigenenergies of the low-lying scattering states of the relative motion of the system based on a set of many-nucleon structure calculations.
With the eigenenergies, we extrapolate the desired elastic scattering phase shift based on the MERE formula to the limit of vanishing external HO potential.
We propose to compute the eigenenergies of relevant many-nucleon systems on quantum computers via efficient quantum eigensolvers, where such calculations can be intractable with classical computers.
Subsequent calculations (i.e., to extrapolate the phase shift based on the MERE formula) are easy on a classical computer but hard on quantum computers; they are therefore processed on classical computers.

We illustrate our framework with two simple toy problems, the neutron-proton ($np$) scattering in the $0^+$ channel and the neutron-$^4$He ($n\alpha$) scattering in the $(1/2)^+$ channel. 
We adopt the rodeo algorithm \cite{PhysRevLett.127.040505,qian2021demonstration,beelindgren2022rodeo} to solve the eigenenergies of the discretized low-lying scattering states of the relative motions of the scattering systems that are confined in the HO potentials, where the quantum computations are simulated with the IBM Qiskit package \cite{Qiskit}. We also present the extrapolation schemes to obtain the phase shifts based on these eigenenergies. 
Being limited in the resources required for general many-body calculations on quantum computers, we take the nuclei in our toy problems to be structureless particles. We aim for complex many-body applications with our framework on future fault-tolerant quantum computers \cite{shor1997faulttolerant,preskill1997faulttolerant,Gottesman_1998}.

The arrangement of this paper is as follows. We present the theory of our framework in Sec. \ref{sec:theory_part}, which includes the MERE formula, the elastic scattering between two nuclei, and the rodeo algorithm. A brief summary of our framework is also provided at the end of Sec. \ref{sec:theory_part}. We illustrate our framework with two toy problems in Sec. \ref{sec:toyModel}. We conclude in Sec. \ref{sec:conclusion_and_outlook}, where we also present the outlook. We provide supplementary materials in the Appendices.

\section{Theory}
\label{sec:theory_part}

We aim to solve the phase shift of the elastic scattering between two nuclei $\mathcal{X}$ (of $B$ constituent nucleons) and $\mathcal{Y}$ (of $A-B$ constituent nucleons) in an uncoupled channel at low energies. For such cases, we assume both nuclei are well bound and the nuclei remain at the kinematically allowed ground states during the scattering. In this section, we introduce the elements of our framework. We summarize the framework is at the end of this section.

\subsection{The MERE formula}
\label{sec:MERE_formula_section}

The Hamiltonian of two point particles in their center of mass (CM) frame reads\footnote{We adopt natural units and take $\hbar=c=1$ in this work.}
\begin{equation}
	H_{\rm sc} = T_{\rm rel} + V_{\rm int} ,
	\label{eq:original_H}
\end{equation}
where $ T_{\rm rel} ={\vec{p}^2}/({2 \mu }) $ denotes the kinetic energy of the relative motion of the two particles, with $ \vec{p} $ being the relative momentum and $ \mu $ being the reduced mass. $ V_{\rm int} $ denotes the interaction between the particles, which is assumed to be of finite range. The wave function of the scattering system and the phase shift can be solved based on $H_{\rm sc} $ with the scattering boundary condition \cite{goldberger2004collision,bhaduri1975structure}.

Complementary to the traditional approaches described above, one can also solve the elastic scattering phase shift based on the energy eigenvalues of the scattering system that is subjected to some external confinement. As a textbook problem in nuclear physics, the scattering phase shift of two nucleons confined in a spherical region can be obtained from the energy eigenvalues of the system by solving the Schr\"odinger equation with the Dirichlet boundary condition [see, e.g., Ref. \cite{goldberger2004collision,bhaduri1975structure}].

Following this idea, one considers the confining potential to be the harmonic oscillator (HO) potential $ V_{\rm HO} (\omega) = (1/2) \mu \omega ^2 r^2 $, where $\omega $ being the oscillator strength (referred to as the ``trap strength" in the following) and $r$ being the distance between the mass centers of the two particles. The external confining potential acts only on the relative degrees of freedom of the scattering system. The Hamiltonian of the scattering system that is subjected to the HO potential becomes 
\begin{equation}
	H (\omega) = H_{\rm sc} + V_{\rm HO}(\omega) =  T_{\rm rel} + V_{\rm int} + V_{\rm HO}(\omega).
	\label{eq:confined_system}
\end{equation}
The ``modified" phase shift $ \delta _{l,\omega} $ of the scattering system confined in the external HO potential in an uncoupled channel is related to the energy eigenvalues of $H (\omega) $ according to the MERE formula \cite{busch1998two, Stetcu_2007_PRA, Suzuki_2009, STETCU20101644, PhysRevC.82.034003,Blume_2012,PhysRevC.101.051602}
\begin{equation}
	p^{2l+1}\cot \delta_{l,\omega}(p)=(-1)^{l+1}(4 \mu \omega)^{l+\frac{1}{2}}\frac{\Gamma\left( \frac{2l+3}{4}-\frac{\epsilon}{2} \right)}{\Gamma\left( \frac{1-2l}{4}-\frac{\epsilon}{2} \right)}, 
	\label{eq:MERE_formula}
\end{equation}
where $\epsilon = E / \omega $ with $ E = {p}^2/(2\mu) $ denoting the eigenenergy of $H (\omega)$. $\Gamma (\cdot )$ denotes the Gamma function \cite{arfken2013mathematical}. $l$ denotes the orbital angular momentum of the relative motion, which also labels the scattering channel. 
We provide the derivation of Eq. \eqref{eq:MERE_formula} in Appendix \ref{sec:MERE_formula_derivation} for completeness.

We remark that the MERE formula holds in the limit of the zero-range interactions, and is valid up to the inelastic threshold of the uncoupled scattering channel.\footnote{The MERE formula [Eq. \eqref{eq:MERE_formula}] can be generalized, in principle, to the coupled-channel cases.} The confining HO potential $V_{\rm HO}(\omega)$ should be weak on the distance scale of the inter-nucleus interactions. Being non-zero everywhere except at the origin, $V_{\rm HO}(\omega)$ then produces appropriately small modifications to the interior of the nuclear wave function, while it modifies the boundary condition outside the interaction range and discretizes the continuum states of the otherwise free scattering. The scattering phase shift is also modified as the outer range of the interaction now differs from that in the free space. The modified scattering phase shift is dependent on the energy eigenvalue $E$ of the confined scattering system. Based on the knowledge of the spectrum of the confined system, one can extract the continuum scattering amplitude up to finite-range corrections \cite{PhysRevC.82.034003,PhysRevC.101.051602}.

\subsection{Elastic scattering of two nuclei}
\label{sec:many_nucleon_scattering}

As mentioned, we consider the elastic scattering between two well-bound nuclei $\mathcal{X}$ (of $B$ constituent nucleons) and $\mathcal{Y}$ (of $A-B$ constituent nucleons).  We apply a weak HO potential (of strength $\omega$) to the scattering system $\mathcal{X}+\mathcal{Y}$, where the HO potential acts to each pair of nucleons in the combined system. The Hamiltonian of the intrinsic motion of the resulting $A$-nucleon system reads
\begin{align}
	H_A(\omega ) = \underbrace{ \frac{1}{2Am} \sum _{i<j}^A (\vec{p}_i - \vec{p}_j)^2 + \sum _{i<j}^A V_{ij} }_{ H_{{\rm sc}, A} } + \underbrace{ \frac{1}{2A} m \omega ^2 \sum _{i<j}^A (\vec{r}_i - \vec{r}_j)^2 }_{\mathcal{V}_{\rm HO}^{\rm rel}(\omega)} ,
	\label{eq:many_nucleon_Hamiltonian_in_trap}
\end{align}
where $H_{\rm sc, A}$ denotes the Hamiltonian of the $A$-nucleon system \cite{Navratil:2000ww, Navratil:2000gs, Barrett:2013nh} that consists of all the nucleons in the colliding nuclei. $m$ denotes the nucleon mass. For simplicity, we take the neutron and proton to be of equal mass. $ V_{ij} $ denotes the inter-nucleon interaction between the $i^{\rm th}$ and $j^{\rm th}$ nucleons.\footnote{We omit the contributions from many-nucleon interactions terms in this work, the inclusion of which is straightforward but involved.} $\mathcal{V}_{\rm HO}^{\rm rel} (\omega)$ denotes the weak HO potential of oscillator strength $\omega $ that acts on pairwise nucleons. $\vec{p}_i$ and $\vec{r}_i$ denote the momentum and position of the $i^{\rm th}$ nucleon, respectively.

We consider the low-lying scattering states of the $A$-nucleon system; with the HO potential, these states are discretized. Since $\mathcal{X}$ and $\mathcal{Y}$ are taken as well bound nuclei and the HO potential is sufficiently weak, we argue according to the kinematics that $\mathcal{X}$ and $\mathcal{Y}$ are in their respective ground states as the effect of the net nuclear environments generates small perturbations to the energy levels compared to their large excitation energies. Correspondingly, we have in mind that the computational advantage appears when we adopt an ans\"atz wave function for the low-lying states of $H_A(\omega ) $ to be the tensor product of the ground state wave function of the $\mathcal{X}'$ cluster (corresponding to $\mathcal{X}$ in the nuclear environment), the ground state wave function of the $\mathcal{Y}'$ cluster (corresponding to $\mathcal{Y}$ in the nuclear environment), and the wave function of the relative motion of the clusters with proper antisymmetrization. We imagine that this can be easily generalized to include more configurations of the projectile-target system following the resonating group method \cite{TANG1978167} and the cluster decomposition method \cite{Polyzou:1978wp, Martin:2023wrd}.
Correspondingly, it is convenient to divide the eigenenergies $ \{ E_{{\rm tot}, i}(\omega) \} $ of the low-lying scattering states of $H_A(\omega )$ into the sum of 1) the ground-state energy $ E_{\mathcal{X}', {\rm gs}} (\omega) $ of the $\mathcal{X}'$ cluster; 2) the ground-state energy $ E_{\mathcal{Y}', {\rm gs}} (\omega) $ of the $\mathcal{Y}'$ cluster; and 3) the energies $ \{ E_{{\rm rel}, i} (\omega) \} $ of the relative motion of the two clusters. That is, we have 
\begin{equation}
	E_{{\rm tot}, i} (\omega ) = E_{{\rm rel}, i} (\omega ) + E_{\mathcal{X}', {\rm gs}} (\omega) + E_{\mathcal{Y}', {\rm gs}} (\omega) .
	\label{eq:energy_decomposition_scheme}
\end{equation}

The energy eigenvalues of the clusters and the total system can be calculated based on the intuition of the asymptotic kinematics of the low-energy scattering states of the confined $A$-nucleon system. In particular, we can sort $H _A(\omega ) $ according to the kinematics and the ans\"atz wave function as [Eq. \eqref{eq:clusterization_2}]
\begin{align}
	H_A(\omega ) = H_{\rm rel} (\omega ) + H_{ \mathcal{X}'} (\omega) + H_{ \mathcal{Y}' } (\omega), 
	\label{eq:cluster_Hamiltonians} 
\end{align}
where the derivations are presented in Appendix \ref{sec:appendix_MBHamiltonian}. $H_{\rm rel} (\omega ) $ corresponds to the intrinsic motion of the two clusters [Eq. \eqref{eq:Ham_HAB_rel}]
\begin{equation}
	H_{\rm rel} (\omega ) = \underbrace{ \frac{1}{2 \mu } \vec{p}^2 }_{T_{\rm rel}} + \underbrace{ \frac{1}{2} \mu \omega ^2 \vec{r}^2 }_{V_{\rm HO}(\omega)} + \underbrace{ \sum _{i=1}^B \sum _{j=B+1}^A V_{ij} }_{ V_{\rm int}} ,
	\label{eq:relative_motion_of_XY}
\end{equation}
with the reduced mass of two clusters being $ \mu = B(A-B)m/A $. $\vec{p}$ is the momentum of the relative motion of the clusters, while $ \vec{r} $ denotes the relative position of the $\mathcal{X}'$ and $\mathcal{Y}'$ clusters. The last term of the above equation denotes the interaction $ V_{\rm int}$ between the two clusters. $ V_{\rm int} $ can be derived from first principles or modeled as phenomenological potentials. 

We remark that Eq. \eqref{eq:relative_motion_of_XY} is in accordance with Eq. \eqref{eq:confined_system}. By direct comparisons, we note that $H_{\rm rel} (\omega)$ [Eq. \eqref{eq:relative_motion_of_XY}] resembles the Hamiltonian $H(\omega)$ [Eq. \eqref{eq:confined_system}] of the relative motion of the $\mathcal{X}$ and $\mathcal{Y}$ nuclei (as if they were two structureless particles) with the additional HO potential and the inter-nucleus interaction. These two Hamiltonians produce the same eigenenergies $\{ E_{{\rm rel}, i} (\omega ) \} $ below the inelastic threshold. 

The Hamiltonian of the cluster $\mathcal{X}'$ is $H_{ \mathcal{X}'} (\omega) $. With the scaled mass $m' = Am/B $ and scaled oscillator strength $\omega ' = B\omega /A$, $ H_{ \mathcal{X}'} (\omega) $ can be written as [Eq. \eqref{eq:Ham_HB}]
\begin{multline}
	H_{ \mathcal{X}'} (\omega ) = 
	\frac{1}{2Bm'}  \sum _{i=1, i<j}^B (\vec{k}_i - \vec{k}_j)^2 + \frac{m' (\omega ') ^2}{2B}  \sum _{i=1, i<j}^B (\vec{s}_i - \vec{s}_j)^2 +  \sum _{i=1, i<j}^B V_{ij} \\
	+ \frac{A-B}{B} \left[ \frac{1}{2m'} \sum _{i=1}^B \vec{k}_i^2 + \frac{1}{2} m' (\omega ') ^2 \sum _{i=1}^B \vec{s}_i^2 \right] ,
	\label{eq:cluster_in_trap_X}
\end{multline}
where $\vec{s}_i$ denotes the position of the $i^{\rm th}$ nucleon of the $\mathcal{X}'$ cluster with respect to its mass center. The momentum of the $i^{\rm th}$ nucleon is $\vec{k}_i = m \dot{\vec{s}}_i$. The last term of the above equation denotes the effect of the nuclear environment to the $B$ nucleons in the cluster, where each nucleon behaves as a harmonic oscillator with the scaled mass $m'$ and the scaled oscillator strength $\omega '$. The ground state energy of $H_{ \mathcal{X}'} (\omega )$ is $  E_{\mathcal{X}', {\rm gs}} (\omega)$. 

Similarly, $H_{ \mathcal{Y}'} (\omega) $ denotes the Hamiltonian of the cluster $\mathcal{Y}'$. Taking $m'' = Am/(A-B) $ and $\omega '' = (A-B)\omega /A$, $H_{ \mathcal{Y}'} (\omega) $ can be written as [Eq. \eqref{eq:Ham_HA_B}]
\begin{multline}
	H_{ \mathcal{Y}'} (\omega ) =
	\frac{1}{2(A-B)m''} \sum _{i=B+1, i<j} ^A (\vec{q}_i - \vec{q}_j)^2 + \frac{1}{2(A-B)} m'' (\omega'') ^2 \sum _{i=B+1, i<j} ^A (\vec{t}_i - \vec{t}_j)^2 + \sum _{i=B+1, i<j} ^A  V_{ij} \\
	+ \frac{B}{A-B} \left[ \frac{1}{2m''} \sum _{i=B+1}^A \vec{q}^2_i 
	+  \frac{1}{2} m'' (\omega '') ^2 \sum _{i=B+1}^A \vec{t}^2_i \right] ,
	\label{eq:cluster_in_trap_Y}
\end{multline}
where $\vec{t}_i$ denotes the position of the $i^{\rm th}$ nucleon of the $\mathcal{Y}'$ cluster with respect to the cluster mass center. The momentum $\vec{q}_i = m \dot{\vec{t}}_i$ is conjugate to $\vec{t}_i$. The last term in Eq. \eqref{eq:cluster_in_trap_Y} results from the effect of the nuclear environment to the $(A-B)$ nucleons in the cluster. The ground state energy of $ H_{ \mathcal{Y}'} (\omega ) $ is $ E_{\mathcal{Y}', {\rm gs}}(\omega) $. 

In the extreme case where both nuclei are structureless, there is no contribution from the intrinsic motion of the clusters to the total Hamiltonian $H_A (\omega ) $. In this limit, the total Hamiltonian $H_A(\omega ) $ [Eq. \eqref{eq:cluster_Hamiltonians}] of the intrinsic motion of the $A$-nucleon system is simply $ H_{\rm rel} (\omega ) $, as discussed in the Appendix \ref{sec:appendix_MBHamiltonian}. 

We can infer the eigenenergies $\{ E_{{\rm rel}, i} (\omega ) \} $ of $ H_{\rm rel} (\omega ) $ according to Eq. \eqref{eq:energy_decomposition_scheme}, with the energy eigenvalues $\{  E_{{\rm tot}, i} (\omega ) \} $, $E_{\mathcal{X}', {\rm gs}} (\omega) $, and $ E_{\mathcal{Y}', {\rm gs}} (\omega) $ solved from $H_A(\omega )$, $H_{ \mathcal{X}'} (\omega)$, and $H_{ \mathcal{Y}'} (\omega)$, respectively.\footnote{In practical numerical calculations for the low-lying spectra of the Hamiltonians $ H_A(\omega ) , \  H_{ \mathcal{X}'} (\omega)$ , and $ H_{ \mathcal{Y}'} (\omega )  $, additional constraint terms \cite{Lipkin:1958,Gloeckner:1974sst} are helpful to regulate the spectra, e.g., by pruning the CM excitations and by specifying the total angular momentum of the many-nucleon states. Such terms are also helpful to improve the convergence rate of the corresponding many-body calculations. These discussions are included in Appendix \ref{sec:LL_terms_MN_Ham}.} 
With the knowledge of the energy eigenvalues $\{ E_{{\rm rel}, i} (\omega ) \} $ of the relative motion of the colliding nuclei $\mathcal{X}$ and $\mathcal{Y}$, we can extract the corresponding phase shift of the elastic scattering in the uncoupled channel utilizing the MERE formula.

\subsection{Spectral solution via quantum eigensolvers}
\label{sec:QuantumEigenSolver}

We discuss how we can extrapolate the phase shift of the elastic scattering between two well-bound nuclei $\mathcal{X}$ and $\mathcal{Y}$ utilizing the MERE formula [Eq. \eqref{eq:MERE_formula}] based on the low-lying energy eigenvalues of $H_{\rm rel} (\omega ) $, which can be calculated from those eigenenergies of $ H_A(\omega ) $, $H_{ \mathcal{X}'} (\omega ) $, and $H_{ \mathcal{Y}'} (\omega ) $ according to Eq. \eqref{eq:energy_decomposition_scheme}. These energy eigenvalues can be obtained by precision {\it ab initio} calculations based on realistic inter-nucleon interactions utilizing the quantum many-body framework that respects all the known symmetries of the nuclear systems under investigation. However, the quantum many-body calculations are known to be computationally difficult \cite{Barrett:2013nh,Maris:2012du}. We resort to future quantum computing techniques for such spectral solutions of the many-nucleon Hamiltonians.

For our exploratory purposes, we show the utilization of quantum computing in our framework for the structure calculations of the $\mathcal{N}$-nucleon systems based on the Hamiltonian $ H_{\mathcal{N}}(\omega ) \in \{ H_A (\omega ),\ H_{ \mathcal{X}'} (\omega), \  H_{ \mathcal{Y}'}(\omega) \}$ with $\mathcal{N} \in \{A,\ B,\ A-B \}$. In particular, we elect the rodeo algorithm \cite{PhysRevLett.127.040505, qian2021demonstration, beelindgren2022rodeo} for the demonstration of solving the energy eigenvalues in this work, while other quantum eigensolvers may also take place in future works.

\begin{figure*}[!ht] 
	\centering
	\includegraphics[scale=0.4]{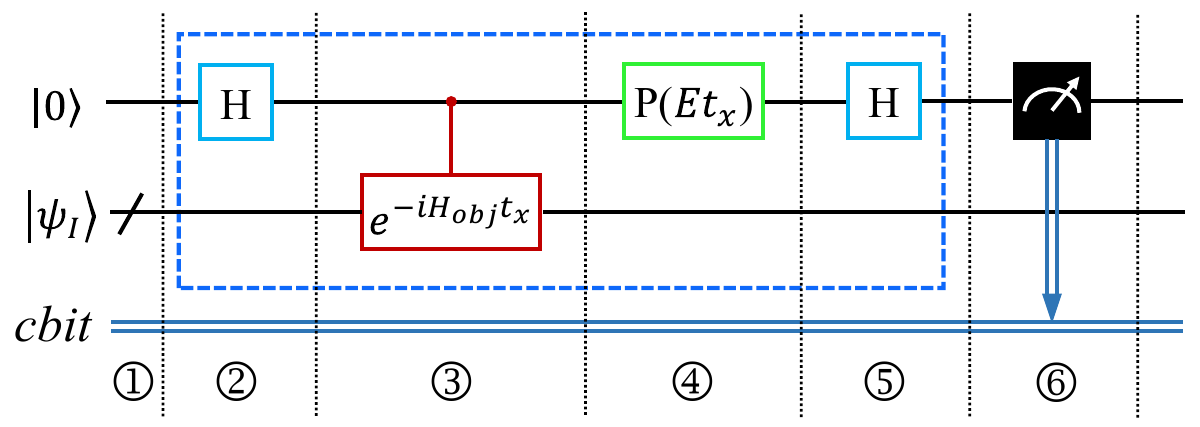}
	\caption{(color online) 
		Demonstration circuit of the rodeo algorithm. The top line denotes the ancilla qubit. The middle slashed line (representing a set of qubits) denotes system register that encodes the state of the system. The bottom double lines (marked as ``{\it cbit}") denote a classical bit that records the result of the measurement (black). ${\rm H}$ (light blue) denotes the Hadamard gate. $P(Et_x)$ (green) denotes the phase gate with two variable $E$ and $t_x$. The controlled rotational gate (red) is constructed based on
		the time-evolution unitary $e^{-i H_{\rm obj} t_x}$ with the objective Hamiltonian $H_{\rm obj} $ of which the spectrum is interrogated, and the variable $t_x$. The ancilla qubit is measured and the result recorded by the classical bit. The circuit (with measurement) is referred to as one rodeo cycle in our discussion.
	}
	\label{fig:rodeo_algorithm_one_cycle}
\end{figure*}

The rodeo algorithm is a probabilistic approach to solve the eigenenergies of both the ground and excited states of the objective Hamiltonian $H_{\rm obj}$ under interrogation; it can also prepare the specific eigenstate for the corresponding energy eigenvalue. We illustrate the typical circuit to implement the rodeo algorithm in Fig. \ref{fig:rodeo_algorithm_one_cycle}. Based on the objective Hamiltonian $H_{\rm obj}$, we have the eigenmode decomposition of the input state $\ket{\psi _I}$ as
\begin{equation}
	\ket{\psi_I}=\sum_j c_j\ket{\phi_j} , 
	\label{eq:inputState_implementation}
\end{equation}
where $\ket{\phi_j}$ denotes the $j^{\rm th}$ eigenstate of $H_{\rm obj}$ that corresponds to the eigenenergy $E_j$. $c_j = \langle \phi _j | \psi _I \rangle $ is the amplitude. 
Provided the input state of the ancilla qubit is $\ket{0}$ and that of the system register is $\ket{\psi _I}$, one performs the following operation
\begin{equation}
	\ket{0} \otimes \ket{\psi _I} \mapsto 
	\frac{1}{2}\sum_jc_j\left[ \left(1+e^{i(E-E_j)t_x}\right)\ket{0} \otimes \ket{\phi_j} -\left(1-e^{i(E-E_j)t_x}\right)\ket{1} \otimes \ket{\phi_j} \right] ,
	\label{eq:one_Rodeo_cycle}
\end{equation}
with the action of the circuit in the red-boxed area in Fig. \ref{fig:rodeo_algorithm_one_cycle}. Then, the probability to measure the ancilla qubit to be in the state $\ket{0}$  after the implementation of one rodeo cycle is \cite{PhysRevLett.127.040505}
\begin{align}
	P_1(E, t_x) \equiv & \sum _j  | c_j |^2 P_{1,j}(E, t_x)  \label{eq:probability_for_one_cycle_without_marginalization}  ,
\end{align}
where $P_{1,j}(E, t_x)  $ is the damping factor 
\begin{equation}
	P_{1,j}(E, t_x) \equiv  \cos^2\left[(E-E_j)t_x /2 \right] .
\end{equation}
$ P_{1,j}(E, t_x) \in [0,\ 1 ] $ is expected to peak at $ E = E_j + 2g \pi /t_x $ with $ g $ being an integer. 

We can also construct the circuit of $R$ rodeo cycles by serializing $R$ circuits of one rodeo cycle. In each rodeo cycle, the variable $E$ is the same while $t_x$ can be different. The ancilla qubit is measured at the end of each rodeo cycle and should be set to $\ket{0}$ at the beginning of each cycle, while the system register is not measured throughout the implementation.

The circuit of $R$ rodeo cycles takes the input $\ket{0} \otimes \ket{\psi _I} $ as that of one rodeo cycle. For a set of Gaussian random variables $\{ t_x \}$ with the mean to be 0 and standard deviation to be $\sigma $, it can be shown that, after marginalizing over $\{ t_x \}$, the probability of obtaining only the $\ket{0}$ state in all of the $R$ consecutive measurements is \cite{qian2021demonstration,beelindgren2022rodeo}
\begin{equation}
	P_R(E, \sigma ) \equiv  \sum _j | c_j |^2 P_{R,j}(E, \sigma ) , \label{eq:marginalization_prbability_1} 
\end{equation}
with the damping factor being
\begin{equation}
	P_{R,j}(E, \sigma ) \equiv  \left[ \left[1+e^{-(E-E_j)^2\sigma^2/2} \right]/2 \right]^R . \label{eq:marginalization_prbability_2}
\end{equation}
In general, $ P_{R,j}(E, \sigma ) \in (0, 1]$ decays when $ | E-E_j | $ increases, while it peaks at $E=E_j$ with the width $1/(\sqrt{R}\sigma)$ \cite{beelindgren2022rodeo}. By scanning over the variable $E$ in executing the circuit of $R$ rodeo cycles, one expects that the profile of $P_R(E, \sigma) $ [Eq. \eqref{eq:marginalization_prbability_1}] contains a set of peaks with the peak locations and peak values approximating the eigenenergies $\{ E_j \}$ and the corresponding $\{ { |c_j | }^2 \} $, respectively. 


\subsection{Summary of the framework}
\label{sec:sketch_of_the_framework}

\begin{figure*}[!ht] 
	\centering
	\includegraphics[scale=0.5]{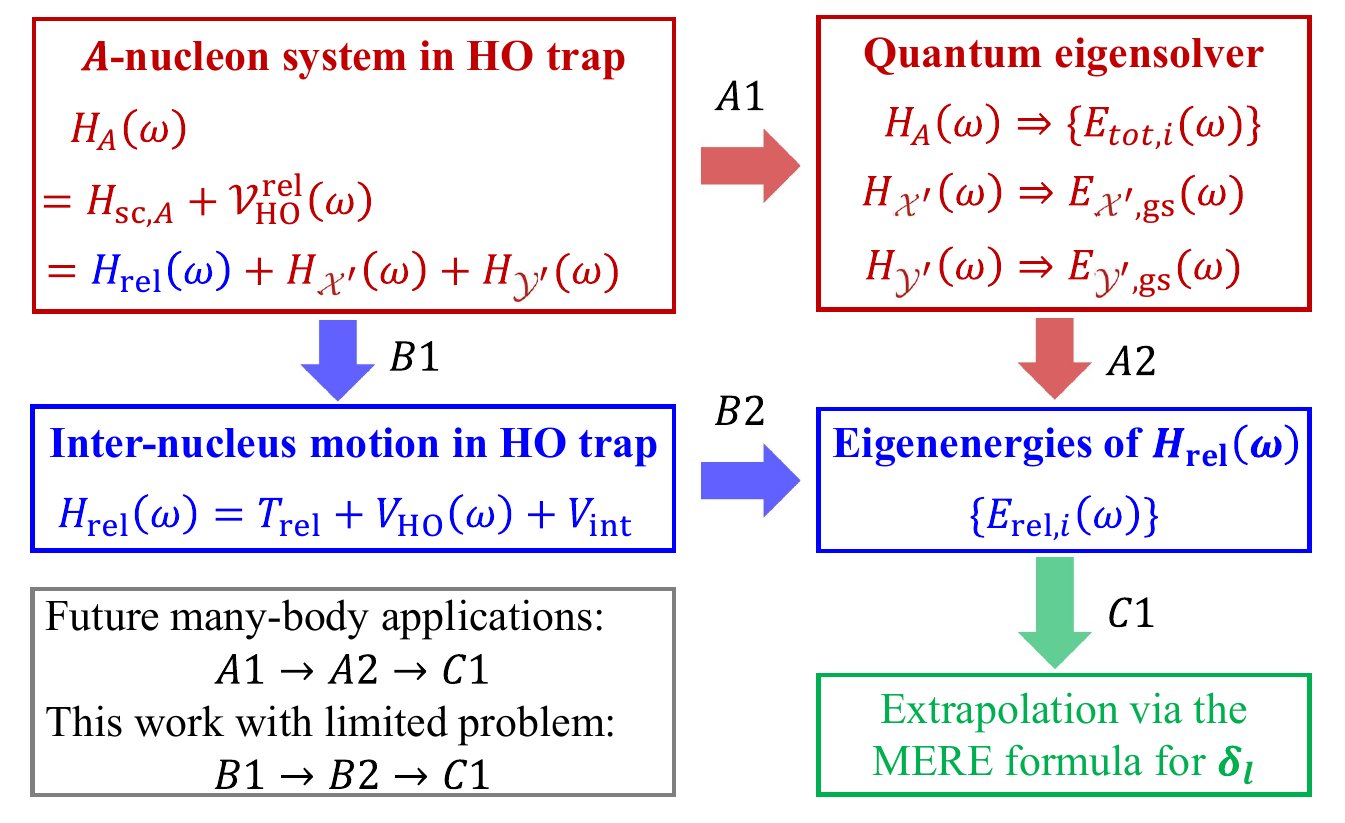}
	\caption{(color online) 
		Workflow of the hybrid quantum-classical approach to solve the phase shift $\delta _l$ of the elastic scattering between two well-bound nuclei in an uncoupled channel. See text for more details.
	}
	\label{fig:framework}
\end{figure*}

We sketch the application of our hybrid quantum-classical framework to a general uncoupled-channel elastic scattering between two well-bound complex nuclei as the route $A1 \rightarrow A2 \rightarrow C1 $ in Fig. \ref{fig:framework}. We discuss in Sec. \ref{sec:many_nucleon_scattering} that the eigenenergies $\{ E_{{\rm rel},i} (\omega) \}$ of the discretized low-lying scattering states of the Hamiltonian $H_{\rm rel} (\omega )$ of the relative motion between the colliding nuclei when confined by the external HO potential $V_{\rm HO}(\omega)$ can be inferred from the spectral solutions $E_{{\rm tot}, i} (\omega )$, $E_{\mathcal{X}', {\rm gs}} (\omega) $, and $ E_{\mathcal{Y}', {\rm gs}} (\omega) $ of the Hamiltonians $H_A(\omega )$, $H_{ \mathcal{X}'} (\omega)$, and $  H_{ \mathcal{Y}' } (\omega) $, respectively. Based on $\{ E_{{\rm rel},i} (\omega) \}$, we can apply the MERE formula and extrapolate to obtain the scattering phase shift. 
Via this route, the spectral solutions of the many-nucleon Hamiltonians, which is hard for the classical computers, can be obtained via quantum computing; on the other hand, the extrapolation for the phase shift, which is expected to be hard for the quantum computers, can be completed on the classical computers.

The route $A1 \rightarrow A2 $ takes the input of the inter-nucleon interactions and infers the inter-nucleus interaction $V_{\rm int}$ based on the inter-nucleon interactions [see Eq. \eqref{eq:relative_motion_of_XY}]
\begin{equation}
	V_{\rm int} = \sum _{i=1}^B \sum _{j=B+1}^A V_{ij}.
\end{equation}
On the other hand, if the form of $V_{\rm int}$ is solved from the inter-nucleon interactions, it can be input to Eq. \eqref{eq:relative_motion_of_XY}. Then, via the route $B1 \rightarrow B2 $, one can solve the energy eigenvalues $\{ E_{{\rm rel},i} (\omega) \}$ of the discretized low-lying scattering states of the Hamiltonian $H_{\rm rel} (\omega )$ [Eq. \eqref{eq:relative_motion_of_XY}] in the uncoupled channel below the inelastic threshold. In principle, the routes $A1 \rightarrow A2 $ and $B1 \rightarrow B2 $ produce the same eigenenergies $\{ E_{{\rm rel},i} (\omega) \}$.

\section{Applications}
\label{sec:toyModel}

In this work, we do not present numerical examples for general applications of our hybrid framework since the spectral solution for a complex many-nucleon system via quantum computing is involving and will be the focus of other works \cite{Du:2023bpw,Du:2024zvr,Liu:2024hmm}. 
Instead, we illustrate our framework with two limited examples: 1) the toy model problem of the $np$ elastic scattering in the $0^+$ channel; and 2) the realistic $n\alpha$ elastic scattering in the $(1/2)^+$ channel. In each problem, the two colliding nuclei are considered as structureless particles, whereas the inter-nucleus interactions $V_{\rm int}$ is modeled as a phenomenological potential for simplicity. With both nuclei being structureless, $H_A(\omega)$ reduces to $H_{\rm rel} (\omega)$ [Eq. \eqref{eq:relative_motion_of_XY}, or, equivalently, Eq. \eqref{eq:confined_system}].

We first present the elastic scattering phase shift of the $np$ system in Sec. \ref{sec:model_problem_of_np}, where we provide 1) the necessary details of the Hamiltonian and the basis representation; 2) the construction of the effective Hamiltonians adopted for further simplifications of the numerical demonstration; 3) our scheme to solve the eigenenergies $\{ E_{{\rm rel},i} (\omega) \}$ of the low-lying discretized scattering states via the quantum eigensolver (elected to be the rodeo algorithm); and 4) the procedures to extract the phase shift based on the MERE formula. The calculation of the $n\alpha$ elastic scattering in Sec. \ref{sec:model_problem_of_nalpha} follows suit.

\subsection{Model problem 1: $np$ scattering}
\label{sec:model_problem_of_np}

We first consider the toy problem of the $np$ elastic scattering in the $0^+$ channel. 
We take the interaction of the $np$ system to be a spherical well potential
\begin{eqnarray}
	 V_{{\rm int},np} = 
	\begin{cases}
		\ -V_0  & \mbox{for} \ r\leq W_0\\
		\ 0     & \mbox{for} \ r > W_0
	\end{cases} , 
	\label{eq:np_spherical_square_well}
\end{eqnarray}
where we use the subscript ``$n p $" to specify the system.
We take the depth of the potential well to be $V_0=48.0002$ MeV and the width of the potential well to be $W_0=1.70134$ fm. We take the neutron mass and the proton mass to be the same $ m=938.919 $ MeV. The reduced mass of the $np$ system is $ \mu _{ np} = 469.460 $ MeV. 
Utilizing the formalism in Ref. \cite{Suzuki_2009}, the effective range and scattering length for the spherical well potential are 5.20 fm and 1.47 fm, respectively.
With these parameters, the $np$ system interacting via the $V_{{\rm int},np}$ contains one shallow bound state with the binding energy $B_0= 2.22$ MeV that reproduces the deuteron binding energy, where the other states are all in the continuum.

For this toy problem, the scattering phase shift of the $l=0$ partial wave admits the analytical solution \cite{wong1998introductory}
\begin{equation}
	\delta _{l,np} =\arctan\left[ \frac{k_0}{k}\tan(k W_0) \right]-k_0 W_0 + Q\pi, \label{eq:np_scattering_phase_shift_analytical}
\end{equation}
where $k_0=\sqrt{2 \mu _{ np} E}$ and $k=\sqrt{2 \mu _{ np} (E+V_0)}$ with $E $ being the scattering energy in the CM frame. $Q$ denotes an arbitrary integer. 

The phase shift $\delta _{l,np} $ can also be obtained via the MERE formula [Eq. \eqref{eq:MERE_formula}] based on the eigenenergies of the discretized scattering states of the Hamiltonian
\begin{equation}
	H_{{\rm rel}, np}(\omega ) =  T_{{\rm rel},np} + V_{{\rm int},np} + V_{{\rm HO},np}(\omega) .
	\label{eq:confined_system_H_np}
\end{equation}
$H_{{\rm rel}, np}(\omega ) $ corresponds to the relative motion of the $np$ scattering system confined in a weak external HO potential $V_{{\rm HO},np}$ with varying oscillator strength $\omega $. 
We present the details of the numerical calculation below, where the results of $\delta _{l,np} $ are shown in Fig. \ref{fig:final_results_np}.

\subsubsection{Basis choice and Hamiltonian matrix}
\label{sec:basis_and_hamiltonian_matrix}

We work in the CM coordinates of the $np$ system and adopt the HO basis to compute the matrix representation of $H_{{\rm rel}, np}(\omega )$. The HO basis  $\ket{nlsJM} $ is specified by a set of quantum numbers: 
1) the radial quantum number $n$; 2) $l$ that labels the orbital angular momentum; 3) $s$ that labels the spin; 4) $J$ that specifies the total angular momentum that is coupled from the orbital angular momentum and the spin; and 5) $M$ that labels the projection of the total angular momentum. 
As for the toy problem of the scattering in the $0^+$ channel, we take $l=0$, $s=0$, $J=0$. We elect $M=0$ in evaluating the Hamiltonian matrix elements (note that $M$ is a good quantum number of the scattering system). 
In practical numerical calculations, we also truncate the HO basis by retaining a set of HO bases $\{ \ket{nlsJM} \}$ with $2n+l \le N_{\rm max}$, where $N_{\rm max} $ denotes the truncation parameter. Besides $N_{\rm max} $, the HO basis set is also specified by the oscillator strength $\Omega $ (referred to as the ``basis" strength, which is not to be confused with the trap strength $\omega $). 

With the HO basis, the matrix elements of $ T_{{\rm rel},np} $ and $  V_{{\rm HO},np} $ admit closed forms [see, e.g., Eqs. (A6) and (A7) in Ref. \cite{Du:2018tce}]. The matrix element of $ V_{{\rm int},np} $ [Eq. \eqref{eq:np_spherical_square_well}] is evaluated via precision numerical integration according to
\begin{align}
	\bra{nlsJM} V_{{\rm int},np} \ket{n'l's'J'M'}&=\delta_{ll'}\delta_{ss'}\delta_{JJ'}\delta_{MM'}\int _0^{\infty}R_{nl}(r) V_{{\rm int},np} R_{n'l'}(r)r^2dr ,
	\label{eq:V_int_in_3DHO}
\end{align}
where the function $R_{nl}(r)$ takes the form 
\begin{equation}
	R_{nl}(r) = \sqrt{\frac{2n!}{b_{np}^3 \Gamma (n+l+3/2) } } \left( \frac{r}{b_{np}} \right)^l \exp \left[ - \frac{r^2}{2b^2_{np}} \right]  L_n^{l+1/2}  \left( \frac{r^2}{b^2_{np}} \right) ,
\end{equation}
with $\Gamma (\cdot )$ being the Gamma function and $L^{\rho}_{n} (\cdot )$ being the generalized Laguerre polynomial \cite{arfken2013mathematical}. The characteristic oscillator length of the HO basis is $b_{\rm np} = \sqrt{1/(\mu _{ np} \Omega ) }$.

\subsubsection{Demonstration with the effective Hamiltonians}
\label{sec:effective_Hamilton}

We elect the trap strength of the external HO potential that confines the $np$ system to be  $ \omega = 3.6, \ 3.7, \ 3.8, \ 3.9, $ and $4.0 $ MeV in our numerical demonstration. We tested that smaller and moderately larger trap strength values also work for the toy problem under consideration. 
We adopt a sufficiently large model space (with $N_{\rm max} = 600 $ at our chosen value of $\Omega = 60$ MeV) for good convergence of the eigenenergies of a set of low-lying discretized scattering states of $H_{{\rm rel}, np}(\omega )$. 

We apply the quantum eigensolver to solve for these eigenenergies. For explanatory purposes, we choose to demonstrate our framework within limited model spaces for simplicity, keeping in mind that the full-scale eigenvalue problems (especially those in future applications with many-nucleon systems) via the quantum eigensolvers can be implemented with straightforward generalizations. 

To this end, we work with the effective Hamiltonians within the limited model space size that are constructed based on the Okubo-Lee-Suzuki method  \cite{Okubo:1954zz,Suzuki:1980yp,Suzuki:1982lbn,Dikmen:2015tla,Vary:2018jxg}. In particular, we design the effective Hamiltonian $H_{{\rm rel}, np}^{\rm eff}(\omega)$ for each $\omega$ value such that it retains the eigenenergies of a set of low-lying eigenstates of the corresponding ``bare" Hamiltonian $H_{{\rm rel}, np}(\omega )$. As in our numerical demonstration, we elect to retain the 1$^{\rm st}$, 2$^{\rm nd}$, 3$^{\rm rd}$, and 4$^{\rm th}$ excited states of $H_{{\rm rel}, np}(\omega )$. Therefore, the matrix of $H_{{\rm rel}, np}^{\rm eff}(\omega)$ is a 4 by 4 matrix.

\subsubsection{Structure calculations via the rodeo algorithm}
\label{sec:rodeo_algorithm_np}

We apply the rodeo algorithm to solve the eigenenergies of $ H_{{\rm rel}, np}^{\rm eff}(\omega) $, which reproduces the eigenenergies of the elected scattering states of $H_{{\rm rel}, np}(\omega )$. The algorithmic details of the rodeo algorithm are reviewed in Sec. \ref{sec:QuantumEigenSolver}. 

An essential part to implement the rodeo algorithm is the circuit construction of the controlled time-evolution unitary $\exp[-i H_{{\rm rel}, np}^{\rm eff}(\omega) t_x]$. In our demonstrations with restricted model space size, we construct the corresponding circuit by the straightforward implementation of the Qiskit package \cite{Qiskit}, where 1) we first express the 4 by 4 time-evolution unitary $\exp[-i H_{{\rm rel}, np}^{\rm eff}(\omega) t_x]$ as a sequence of one- and two-qubit gates in a two-qubit system register; and 2) the control operations determined by the state of the ancilla qubit are then applied to all these gates in sequence. Another important part of the implementation is the choice of the input state. In our demonstrations, we elect the input state to be $\ket{\psi _I} = \frac{1}{2} (1,1,1,1)^T $. This input state can be achieved by two Hadamard gates, each acting on one of the two system qubits. The simulations are performed with the IBM QASM simulator \cite{Qiskit}, where we do not include any simulation noise in this work.

While not being the focus of this work, it is important to note that the efficient circuit construction is critical, especially for future calculations of many-nucleon system, in quantum computing. In particular, the dimension of the Hilbert space in many-body calculations grows rapidly and the straightforward circuit construction scheme adopted in this work results in poor efficiency. Besides, one should prepare proper input states for quantum computing \cite{Lin_2022}. When the input state is of small overlap with a certain eigenstate of which the eigenenergy is under query, the algorithmic implementations can often be inefficient and the outcome can be inaccurate with limited computation resources.
In addition, the noise is also an important feature in quantum computing on near-term quantum hardwares \cite{Preskill2018quantumcomputingin}.
These are critical yet open research topics in quantum computing, which will take their own extensive efforts in future research.

We calculate the eigenenergies of $H_{{\rm rel}, np}^{\rm eff}(\omega)$ via two types of scans over a range of $E$ to solve the eigenenergies. For both scans, we adopt a sufficiently large set of Gaussian random numbers $\{t_x\}$, of which the mean value is $0$ and the standard deviation is $\sigma$ (taken to be 21 MeV$^{-1}$ in this work). 

The first scan is referred to as the ``coarse scan". It returns the approximate values of the eigenenergies. In practice, we apply $R=5$ rodeo cycles in the coarse scan. As the eigenenergies of $H_{{\rm rel}, np}^{\rm eff}(\omega)$ are understood to be positive, we take the range of $E$ to be from $0$ to $\Lambda _m$, where $\Lambda _m $ is larger than the spectral norm of $H_{{\rm rel}, np}^{\rm eff}(\omega)$. The step size of the coarse scan over $E$ is taken to be $\Delta E = 0.1$ MeV. After the coarse scan, we plot the resulting $ P_5(E, \sigma ) $ [Eq. \eqref{eq:marginalization_prbability_1}] as a function of $E$, from which we search for the approximate ranges of the eigenenergies in terms of peak locations. 

A subsequent ``fine scan" is applied to these ranges of the eigenenergies found in the coarse scan for finer resolution. We implement more rodeo cycles ($R=10$) and smaller step size ($\Delta E = 0.001 $ MeV) in the fine scan. From the fine scan, we obtain the probability $ P_{10} (E, \sigma ) $ [Eq. \eqref{eq:marginalization_prbability_1}] as a function of $E$ for each scanned range. We fit each peak of the profile $ P_{10} (E, \sigma ) $ with the function 
\begin{equation}
	f(E) = K_j \left[  \frac{1+e^{-(E-E_j)^2\sigma^2/2}}{2}  \right]^{R} .
	\label{eq:profile_func}
\end{equation}
to obtain the peak center $E_j$, which corresponds to the eigenenergy, and the constant $ K_j $ that corresponds to the overlap $ |c_j |^2$ in Eq. \eqref{eq:marginalization_prbability_1}. 

\begin{figure}[ht]
	\centering
	\includegraphics[scale=0.35]{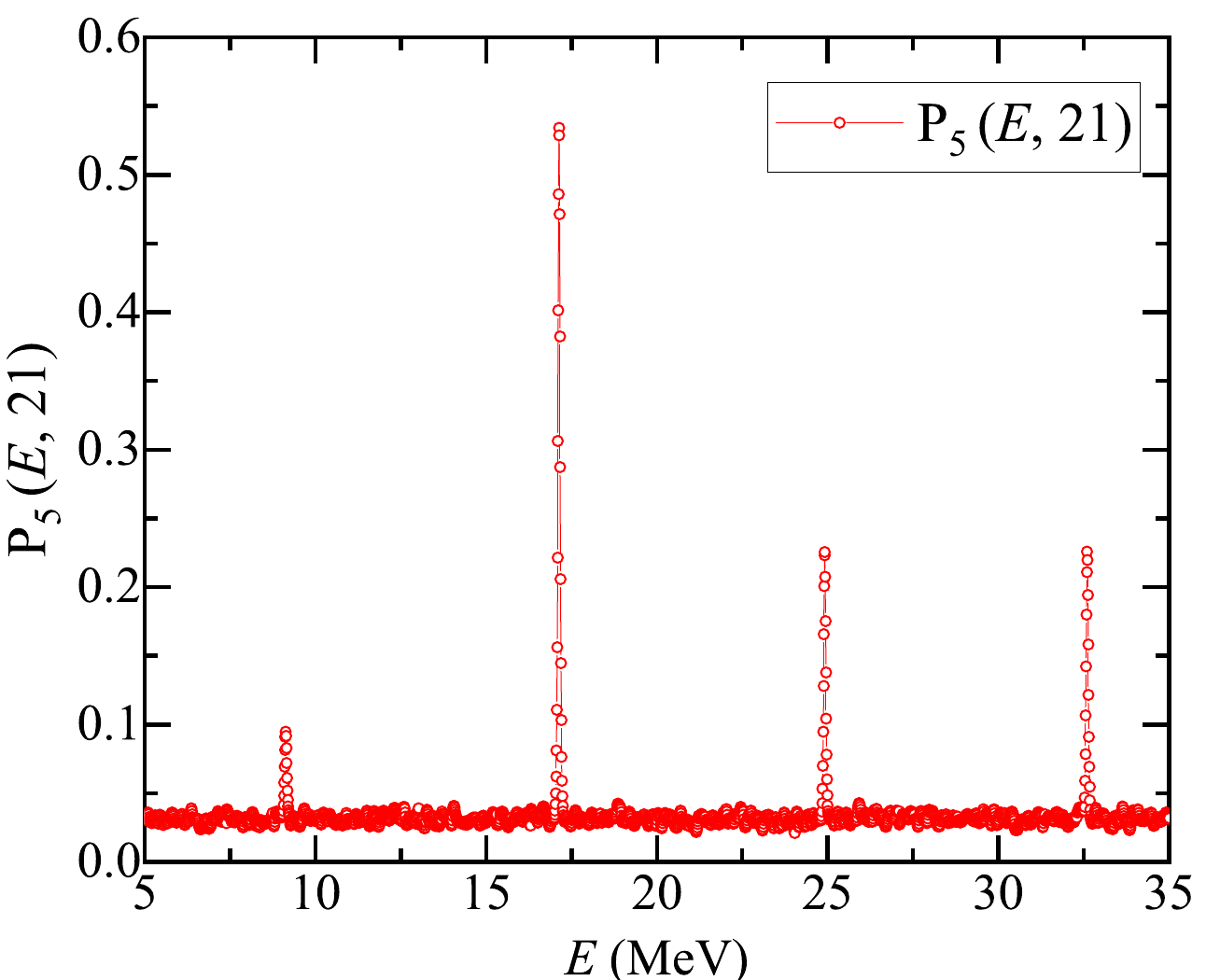}
	\caption{(color online) The coarse-scan results $ P_5(E, \sigma =21 \ {\rm MeV}^{-1} ) $ as a function of the input parameter $E$ for the Hamiltonian $H_{{\rm rel}, np}^{\rm eff}(\omega=3.7\ {\rm MeV})$. The circles denote the success probabilities obtained from the coarse scan over the parameter $E$. These circles are joined by lines to guide the eye.
	}
	\label{fig:37_coarse_scan}
\end{figure}

As an example, we implement the above procedures to obtain the eigenenergies of the Hamiltonian $H_{{\rm rel}, np}^{\rm eff}(\omega)$ with $\omega =3.7$ MeV. Based on the eigenstates of the chosen Hamiltonian, the input state $\ket{\psi _I} = \frac{1}{2} (1,1,1,1)^T$ can be decomposed as
\begin{equation}
	\ket{\psi _I} =-0.260474\ket{\phi_1}+0.724313\ket{\phi_2}-0.451695\ket{\phi_3}-0.451104\ket{\phi_4} ,
\end{equation}
where $\ket{\phi_1}$, $\ket{\phi_2}$, $\ket{\phi_3}$, and $\ket{\phi_4}$ are the eigenstates with the corresponding eigenenergies being 9.12955, 17.1340, 24.9161, and 32.6021 MeV obtained by classical calculations, respectively. With $\ket{\psi _I} $, we present the results obtained via the rodeo algorithm for the example as follows.
\paragraph*{Coarse scan.} We obtain the results of $P_5(E, \sigma ) $ as a function $E$ from the coarse scan. These results are presented in Fig. \ref{fig:37_coarse_scan}. We find that the peaks locate at about 9.0, 17.0, 25.0 and 33.0 MeV. These peak locations provide approximate estimations of the eigenenergies of the Hamiltonian.

\begin{figure}[ht]
	\centering
	\includegraphics[scale=0.35]{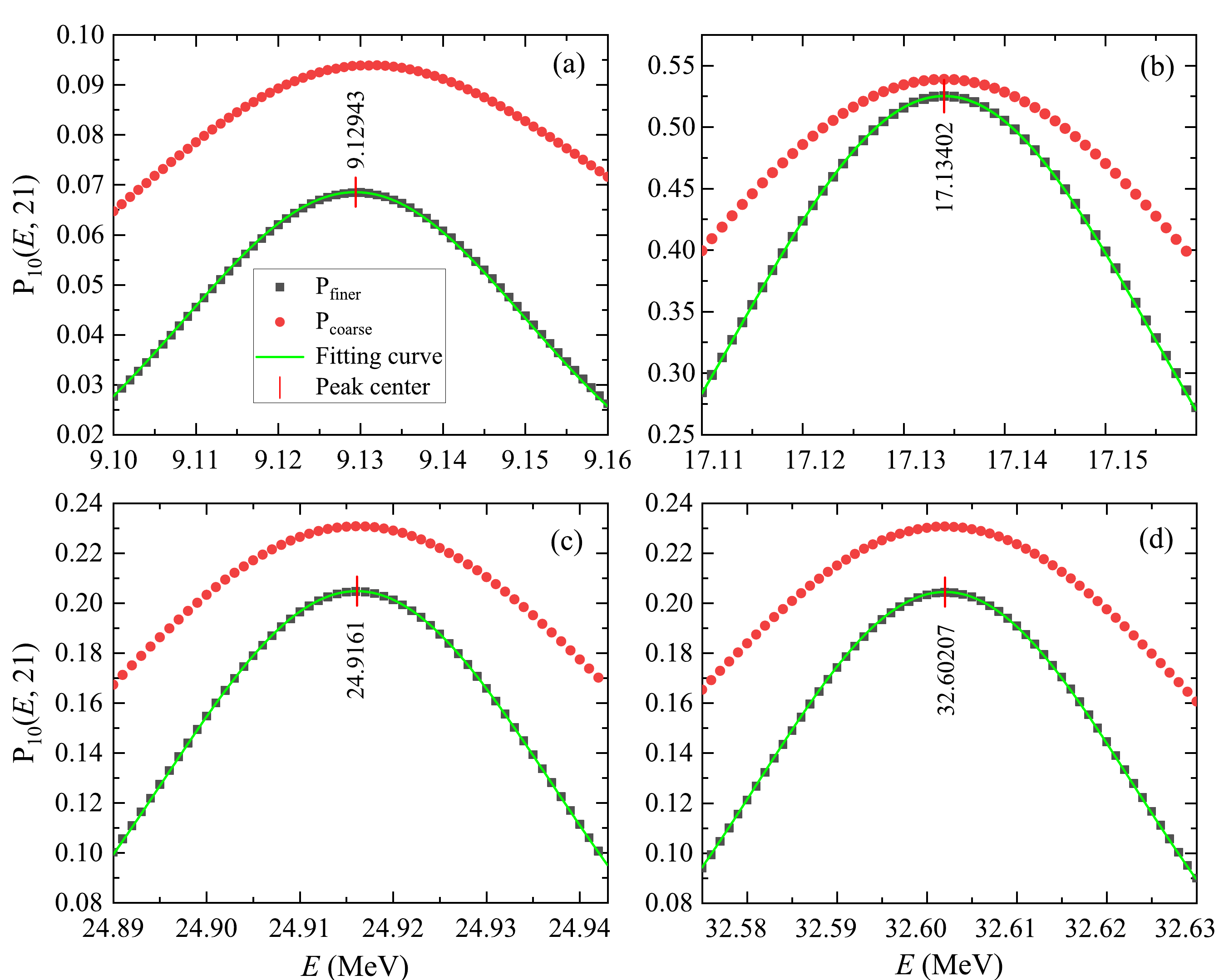}
	\caption{(color online) The fine-scan results $ P_{10} (E, \sigma ) $ as a function of $E$ for the Hamiltonian $H_{{\rm rel}, np}^{\rm eff}(\omega=3.7\ {\rm MeV})$. The fine-scan results are presented in black.
	For direct comparisons, the coarse scan results (red) are computed with the same step size $\Delta E = 0.001 $ MeV as that of the fine scan.
	The fine-scan results are fitted according to Eq. \eqref{eq:profile_func}, with the fitting curves displayed as green lines. The peak centers (red dashes with values) are extracted from the fittings.  
	Panels (a), (b), (c), and (d) are ordered according to the magnitudes of $E$.
	}
	\label{fig:37_all_finer_scan}
\end{figure}

\paragraph*{Fine scan.} Successive fine scan is performed over these approximate ranges where the peaks reside. The fine scan results $ P_{10} (E, \sigma ) $ are presented as a function of $E$ in Fig. \ref{fig:37_all_finer_scan}.
We fit the peaks according to Eq. \eqref{eq:profile_func} and extract the peak centers, of which the central values present our solutions to the eigenenergies of $H_{{\rm rel}, np}^{\rm eff}(\omega=3.7\ {\rm MeV})$ based on the rodeo algorithm. 

We apply the same approach to solve the eigenenergies of the Hamiltonians of the $np$ scattering system confined in the HO trap with strengths $\omega = 3.6, \ 3.8, \ 3.9,\ $and$\ 4.0 $ MeV via respective effective Hamiltonians adopted for numerical simplicity in our illustration. 
In Table \ref{table:classical_and_numerical_energy_eigenvalues_of_np}, we present the eigenenergies of the Hamiltonian $H_{{\rm rel}, np}^{\rm eff}(\omega )$ solved via the rodeo algorithm for all the elected $\omega$ values. We recall that these eigenenergies are also the lowest four scattering states of the corresponding bare Hamiltonian $H_{{\rm rel}, np}(\omega )$ according to our construction of the effective Hamiltonian. 

For comparison, we also present the eigenenergies of the lowest four scattering states of $H_{{\rm rel}, np}(\omega )$ computed via direct matrix diagonalizations on the classical computers in Table \ref{table:classical_and_numerical_energy_eigenvalues_of_np}. We find that, in general, a good agreement between the results obtained based on the rodeo algorithm and those numerically exact results from the direct matrix diagonalizations, where the agreement is up to the 6$^{\rm th}$ significant figure in most cases. We note that the agreement for the case of $E_1$ is relatively poorer that the other cases for all the $\omega $ values. 
This is due to the small overlap $|\langle \phi _1 | \psi _I \rangle|^2$ between the input state and the first excited state, which may also induce numerical artifacts (e.g., the $E_1$ result based on the coarse scan differs noticeably from that obtained via the fine scan, which can be seen from the shift in the peak centers in the upper left panel of Fig. \ref{fig:37_all_finer_scan}).
We can improve the precision of the solutions via the rodeo algorithm (and hence the agreement) by, e.g., 1) increasing the number of rodeo cycles $R$; and 2) taking the input state that has reasonably large overlap with the desired eigenstate, i.e., $\ket{\phi_1}$ which corresponds to $E_1$ in the current model problem. 

\renewcommand{\arraystretch}{1.5}  

\begin{table}[!ht]  
	\setlength{\tabcolsep}{6pt}
	\centering  
		\caption{Eigenenergies of the lowest four scattering states, $E_j$ ($j=1,\ 2,\ 3,\ $and$ \ 4$), of the $np$ system when confined in the HO potentials with strengths $\omega = 3.6, \ 3.7, \ 3.8, \ 3.9,\ $and$\ 4.0 $ MeV. The results denoted as ``exact" are obtained by diagonalizing the bare Hamiltonian $H_{{\rm rel}, np}(\omega )$, while the results obtained by the rodeo algorithm are denoted as ``rodeo". The eigenenergies are in the units of MeVs. See in the text for more details.}  
		\label{table:classical_and_numerical_energy_eigenvalues_of_np}  
		\begin{tabular}{ccccccccccc}  
			\hline
			\hline
			\multirow{2}{*}{Energy}&  
			\multicolumn{2}{c}{$\omega$ = 3.6 MeV}&\multicolumn{2}{c}{$\omega$ = 3.7 MeV}&\multicolumn{2}{c}{  $\omega$ = 3.8 MeV}&\multicolumn{2}{c}{$\omega$ = 3.9 MeV}&\multicolumn{2}{c}{$\omega$ = 4.0 MeV}\cr
			\cline{2-3}\cline{4-5}\cline{6-7}\cline{8-9}\cline{10-11}
			& exact & rodeo & exact & rodeo & exact & rodeo & exact & rodeo & exact & rodeo \cr  
			\hline  
			$E_1$&8.85547&8.85538 &9.12955&9.12943    &9.40445&9.40419    &9.68013&9.68012   &9.95660&9.95661\cr 
			$E_2$&16.6430&16.6430 &17.1340&17.1340    &17.6259&17.6260    &18.1187&18.1187    &18.6123&18.6122\cr 
			$E_3$&24.2138&24.2137 &24.9161&24.9161    &25.6193&25.6193    &26.3233&26.3233    &27.0283&27.0283\cr 
			$E_4$&31.6913&31.6912 &32.6021&32.6021    &33.5138&33.5137    &34.4264&34.4263    &35.3399&35.3400\cr 
			\hline
			\hline
		\end{tabular}  
\end{table}  

\subsubsection{Phase shift extractions}
\label{sec:extrapolation_procedures_np}

\begin{figure}[ht]
	\centering
	\includegraphics[scale = 0.35]{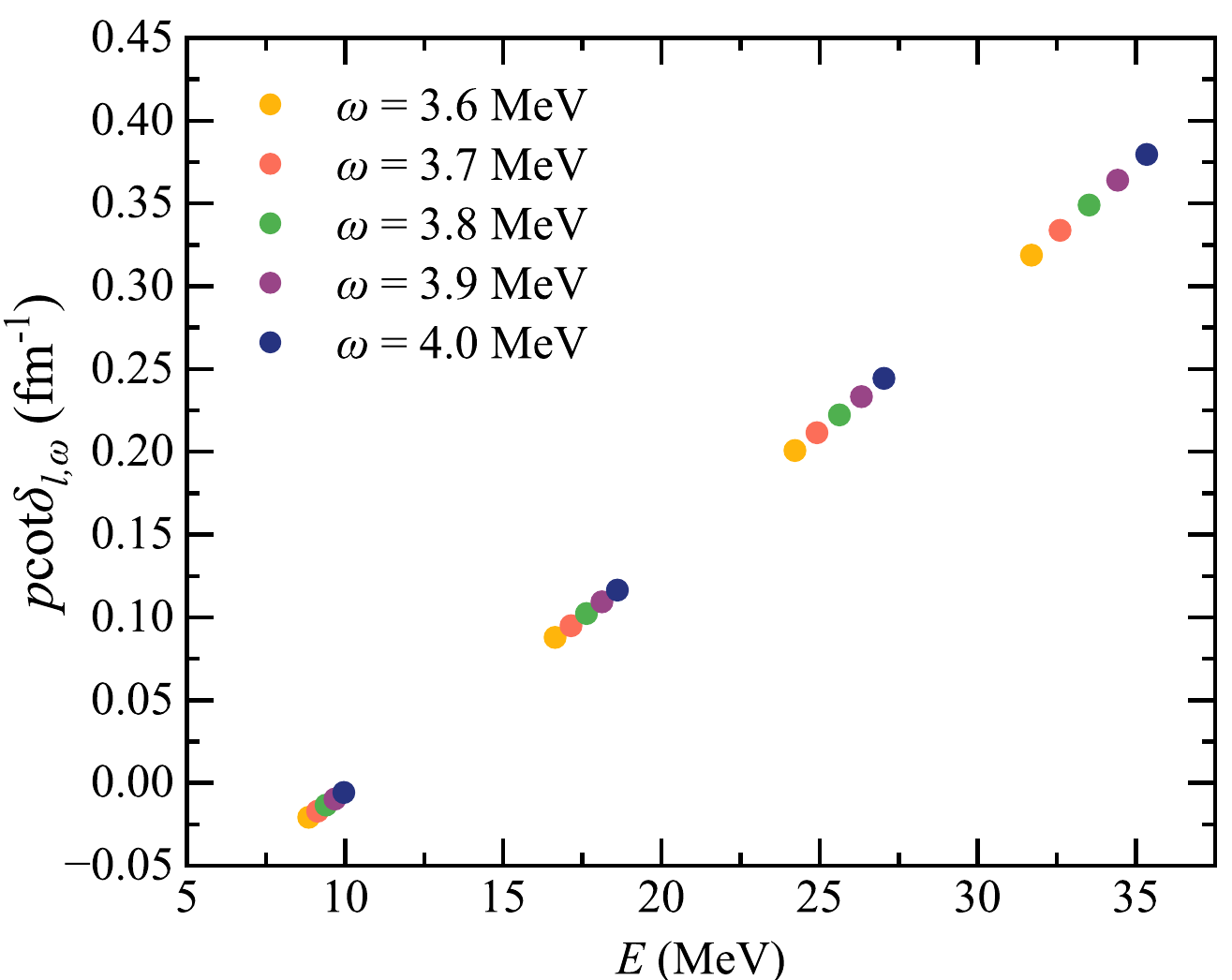}
	\caption{(color online) 
		$p\cot \delta_{l,\omega}(E)$ as a function of the trap strength $\omega$ and the scattering energy $E$ in the CM frame. The results of $p\cot \delta_{l,\omega}(E)$ are extracted according to Eq. \eqref{eq:MERE_formula} from the eigenenergies of the lowest four scattering states of $ H_{{\rm rel}, np}(\omega)$ that are obtained via the rodeo algorithm. 
	}
	\label{fig:E_pcotdelta}
\end{figure}

In this section, we show the procedures to obtain the continuum elastic scattering phase shift from the eigenenergies of the scattering states of $H_{{\rm rel}, np}(\omega )$ obtained via the rodeo algorithm. In particular, based on the data set $\{ E_j (\omega)\}$ in Table \ref{table:classical_and_numerical_energy_eigenvalues_of_np}, we obtain the set $\{E,\ p \cot \delta _{l, \omega } (E) \}$ according to the MERE formula [Eq. \eqref{eq:MERE_formula}], where we recall $l=0$ for the $np$ scattering in the $0^+$ channel. In Fig. \ref{fig:E_pcotdelta}, we present $p\cot \delta_{l,\omega}(E) $ as a function of the scattering energy $E$ and the trap strength $\omega $. 
Based on the set $\{E,\ p \cot \delta _{l, \omega } (E) \}$, we follow the methodology in Ref. \cite{PhysRevC.82.034003} and perform the extrapolation and interpolation in the domains of $E$ and $\omega$ to obtain the continuum phase shift $\delta _{l,\omega = 0} (E) $ at a given $E$.

\begin{figure}[!ht]
	\centering
	\includegraphics[scale = 0.35]{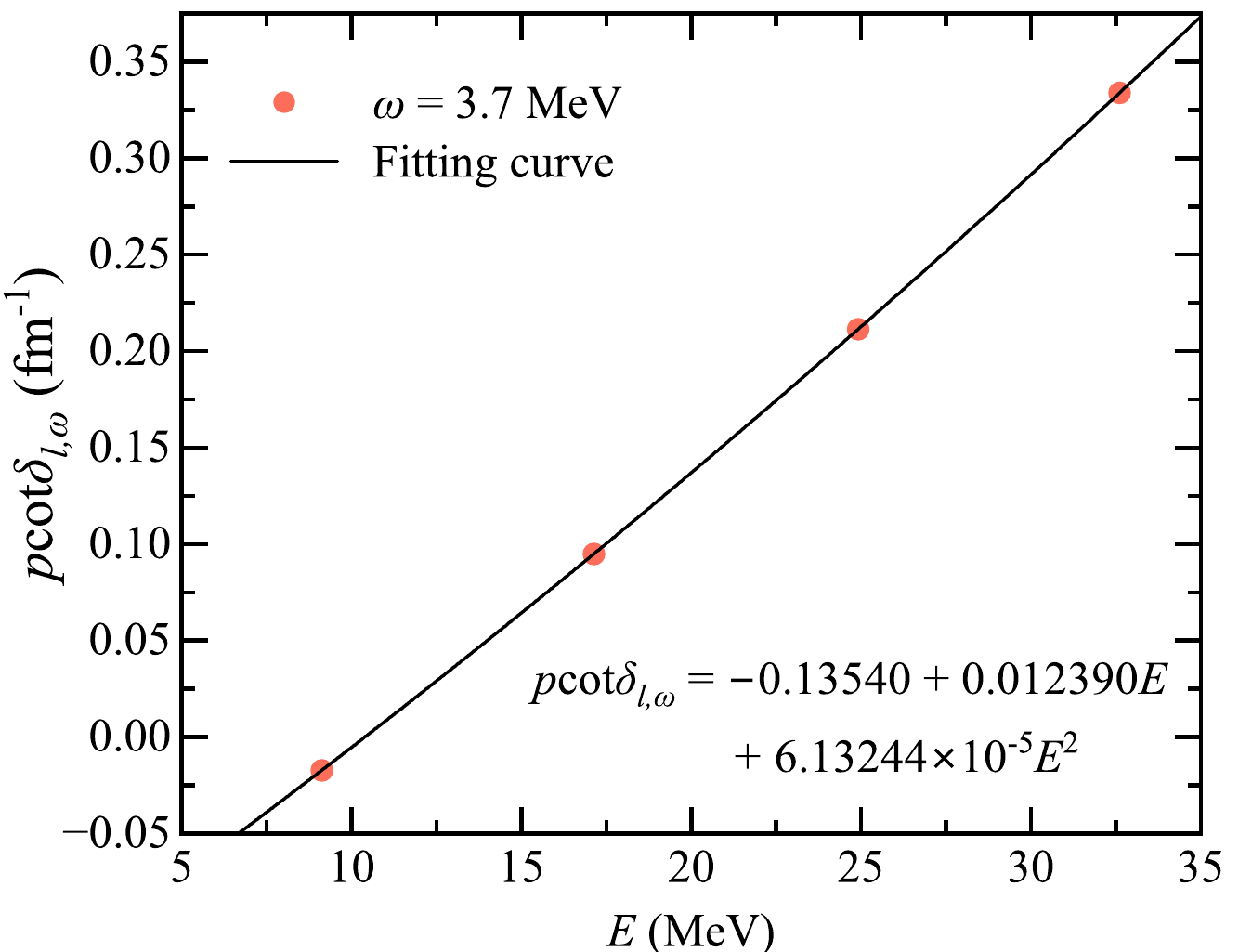}
	\caption{(color online) 
		Extrapolation of $p\cot \delta_{l,\omega}(E)$ as a function of $E$ with fixed $\omega$ (chosen to be 3.7 MeV for demonstration). See the text for details. 
		}
	\label{fig:E_pcotdelta_example_omega3.7MeV}
\end{figure}

The first step is to extrapolate $	p\cot \delta_{l,\omega}(E) $ in the domain of $E$. 
Here $E$ takes the values of the energy eigenvalues of the elected scattering states of $H_{{\rm rel},np} (\omega ) $ in Table \ref{table:classical_and_numerical_energy_eigenvalues_of_np}. These energy eigenvalues are generally different for the Hamiltonian $H_{{\rm rel},np} (\omega ) $ with different $\omega$ values.
For each trap strength $\omega$, we apply the second-order polynomial:
\begin{equation}
	p\cot \delta_{l,\omega}(E)=d_0 + d_1E + d_2E^2 , \label{eq:fourth_order_polynomial_fitting}
\end{equation}
to fit the values of $p \cot \delta _{l, \omega } (E) $, where $d_0$, $d_1$, and $d_2$ are the parameters to be determined. As an example, we present the fitting results as a function of $E$ at fixed $\omega =3.7 $ MeV in Fig. \ref{fig:E_pcotdelta_example_omega3.7MeV}. The fitting procedures of the data set with the other $\omega $ values are the same. 
We remark that the order of $E$ in the polynomial is chosen such that the $\chi ^2 / {\rm DOF}$ of the fitting is minimized, and that the fitting is stable under the change of the polynomial order. In this work, we do not seek to provide a systematic analysis of the uncertainty; such analysis necessitates the variation of the polynomial order, and requires more $E$ values for a given $\omega $ (i.e., a more complete data set that includes more eigenstates). 

At the end of the first step, we obtain five continuous functions of the scattering energy $E$ that are of the type of Eq. \eqref{eq:fourth_order_polynomial_fitting}, where each function is of fixed $\omega \in \{3.6, \ 3.7,\ 3.8, \ 3.9, \ 4.0 \} $. With the knowledge of these continuous functions, we obtain the data set $ \{E,\ p \cot \delta _{l, \omega } (E) \} $ at a chosen $E$ but for five different $\omega$ values.

The second step is to extrapolate the data set $ \{E,\ p \cot \delta _{l, \omega } (E) \} $ in the domain of $\omega $ at fixed $E$. With the function obtained from the extrapolation, we take the limit of $\omega = 0 $, i.e., $p\cot \delta_{l,\omega = 0}(E) $, from which we extract the continuum phase shift $\delta_{l,\omega =0}(E)$ of the elastic $np$ scattering in the $0^+$ channel at the scattering energy $E$.
In practice, we adopt the fitting function of the form \cite{PhysRevC.82.034003}
\begin{equation}
	p\cot\delta_{l,\omega}(E)=a_0+a_1\omega^2 
	\label{eq:linear_fitting} ,
\end{equation}
for the extrapolation. $a_0$ and $a_1$ are the constants to be determined. This fitting function is verified by the fact that the finite-range corrections of the HO trap to the scattering phase shift [Eq. \eqref{eq:MERE_formula}] can be approximated as the expansion \cite{PhysRevC.82.034003}
\begin{equation}
	p\cot \delta_{l,\omega}(E) = p\cot \delta_{l,\omega = 0}(E) + h \omega ^2 + \mathcal{O}(\omega ^3) ,
\end{equation} 
for small $\omega$ such that $\sqrt{ \mu _{\rm np} \omega } \ll   p\cot \delta_{l,\omega}  $.
$ h $ is a constant.

With the extrapolation of the data set $ \{E,\ p \cot \delta _{l, \omega } (E) \} $ according to Eq. \eqref{eq:linear_fitting}, one obtains $a_0$ at the limit of $\omega \rightarrow 0 $. Then, the phase shift at the fixed scattering energy $E$ is 
\begin{equation}
	\delta _{np} (E) \equiv \delta_{l,\omega =0}(E) = \cot ^{-1} \left( a_0/\sqrt{\mu _{\rm np} E } \right). \label{eq:extract_scattering_phase_shift}
\end{equation}

\begin{figure}[!ht]
	\centering
	\includegraphics[scale=0.35]{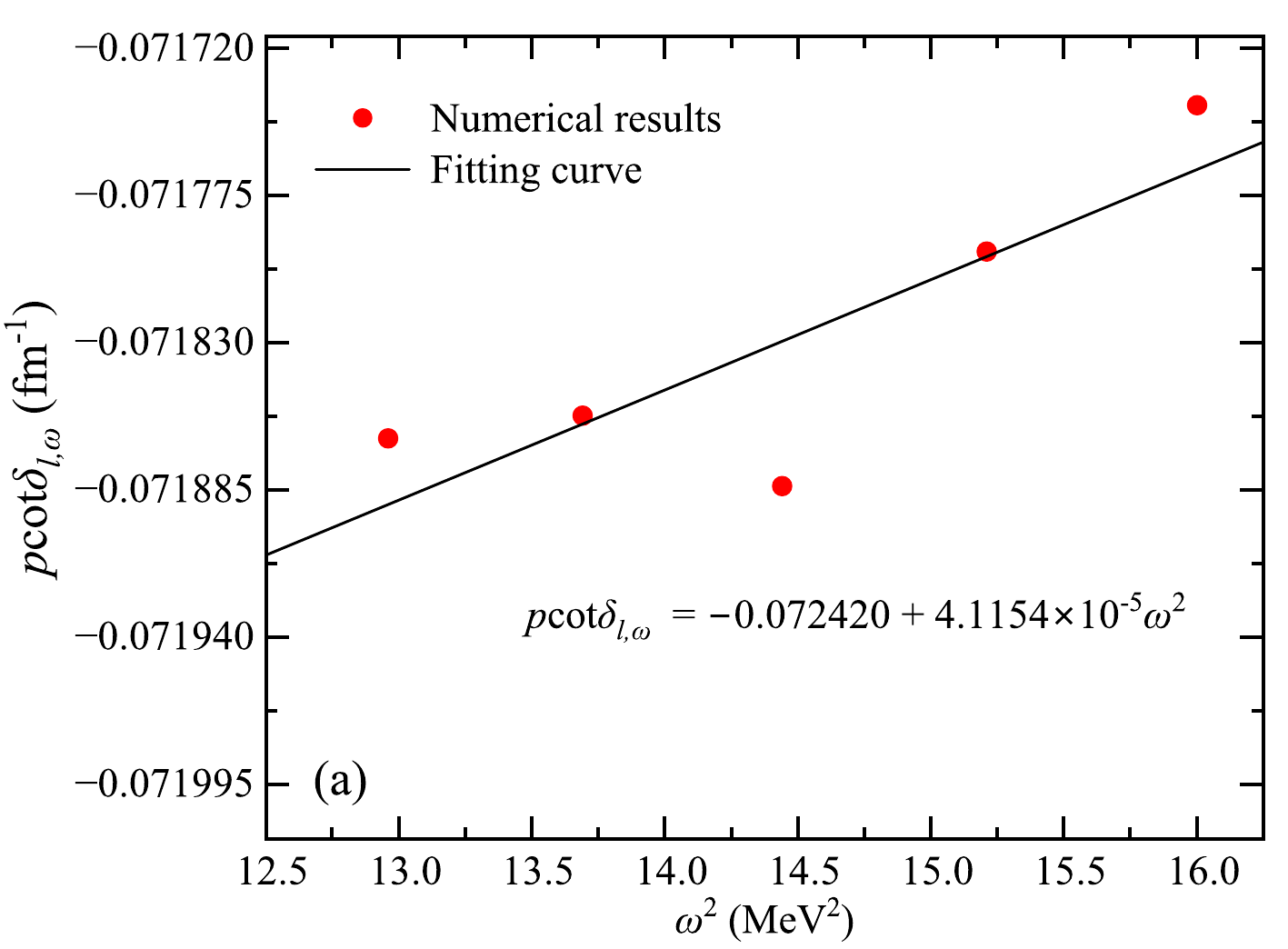}
	\includegraphics[scale=0.35]{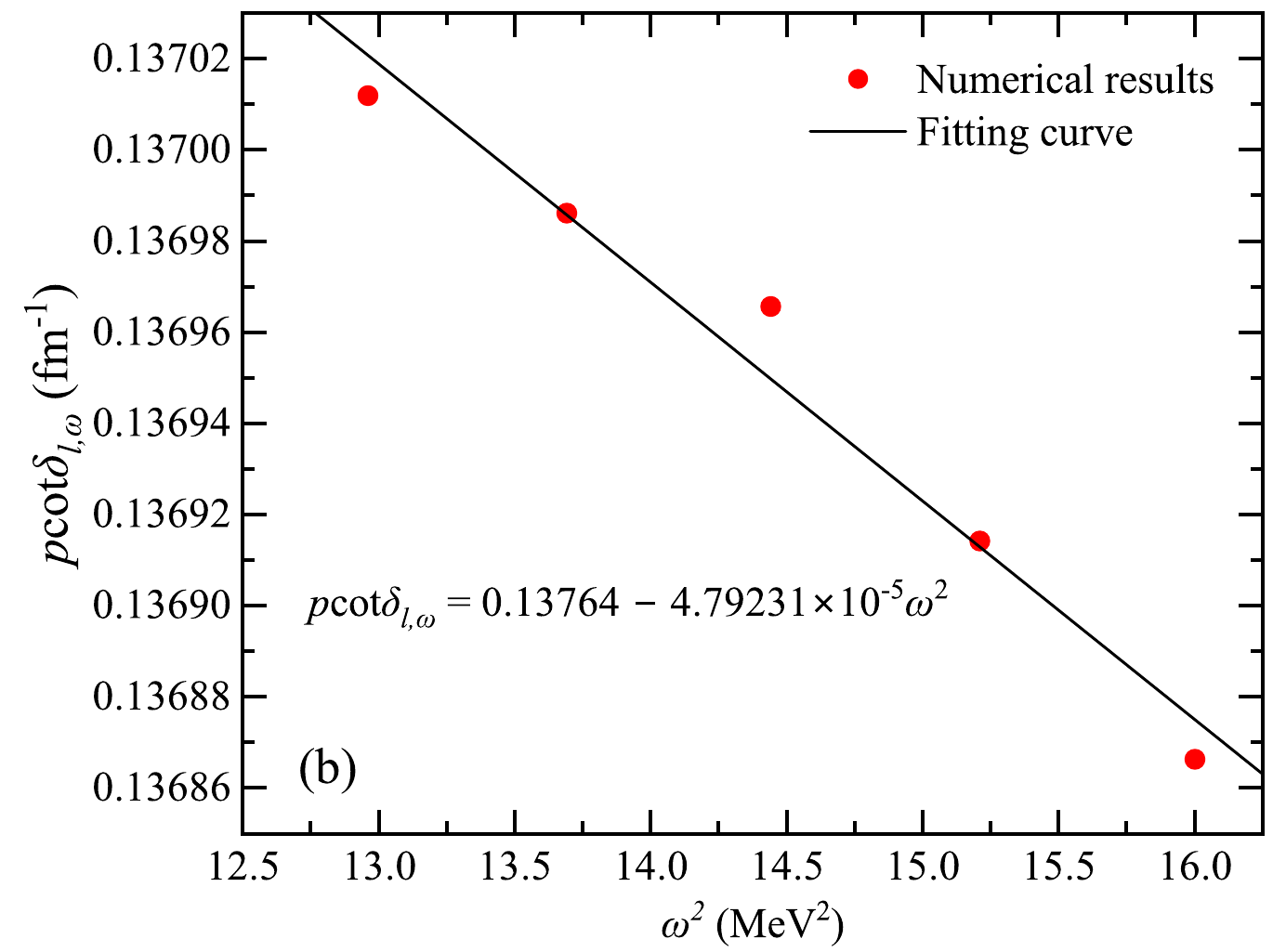}
	\caption{(color online) 
		The extrapolation of $p\cot\delta_{l,\omega}(E)$ as a function of $\omega ^2$ at $E=5 $ MeV (a) and $E=20 $ MeV (b). The solid line in each panel corresponds to the best fit of the functional form of Eq. \eqref{eq:linear_fitting}, whereas the resulting fitting function is shown in each panel.  
	}
	\label{fig:omega-pcotdelta}
\end{figure}

For demonstration purposes, we illustrate the extrapolations of $p\cot \delta_{l,\omega}(E) $ in $\omega $ at the fixed scattering energies $E= 5$ and 20 MeV in Fig. \ref{fig:omega-pcotdelta}. We find that both fitting functions exhibits weak dependence on $\omega ^{2}$. It is noted that the extrapolation for the energy $E=5$ MeV presents larger deviations than the extrapolation for the energy $E=20$ MeV.  This is due to the limitation of our showcase problem: the resulting data set $\{ E_j(\omega) \}$ in Table \ref{table:classical_and_numerical_energy_eigenvalues_of_np}) is sparse in the range of $E\in [8.85547, 35.3400]$ MeV and has no coverage beyond.

\begin{figure}[ht]
	\centering
	\includegraphics[scale=0.40]{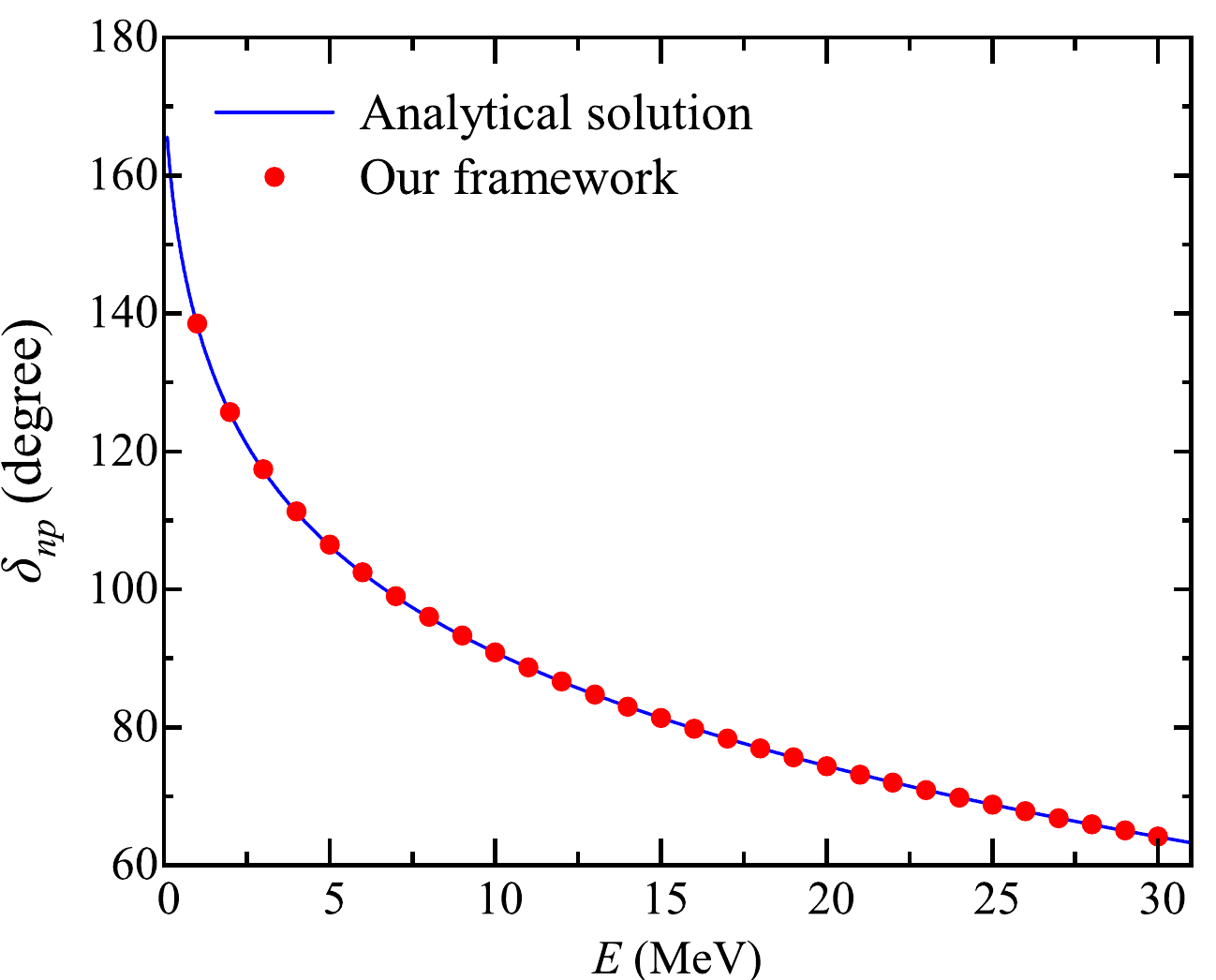}
	\caption{(color online) 
		The $\omega$-extrapolated phase shift $\delta_{np} $ of the elastic $np$ scattering in the $0^+$ channel as a function of the CM scattering energy $E$. 
		The scattering phase shift (red circles) are determined by the interpolation and extrapolation schemes described in the text. These results are computed based on the eigenenergies computed via the rodeo algorithm.
		The solid curve denotes the analytical solution of the $np$ scattering in the $0^+$ channel [Eq. \eqref{eq:np_scattering_phase_shift_analytical}].
	}
	\label{fig:final_results_np}
\end{figure}

With the extrapolation of $\omega \rightarrow 0$ at each scattering energy $E$, we obtain $a_0$ in Eq. \eqref{eq:linear_fitting} and then the phase shift $\delta_{l,\omega =0}(E) $ according to Eq. \eqref{eq:extract_scattering_phase_shift}. In Fig. \ref{fig:final_results_np}, we present our results of the phase shift of the elastic $np$ scattering as a function of the scattering energy $E$ in the CM frame of the system. 
We also compare our results with the analytical solutions [Eq. \eqref{eq:np_scattering_phase_shift_analytical}]. 
We find that our results agree well the analytical solutions, where the deviations are within $0.16\%$.

According to the derivations shown in Appendix \ref{sec:MERE_formula_derivation}, we comment that the MERE formula applies for the cases where the external HO potential is much weaker compared to the strong short-range interaction within the range of the strong interaction. Meanwhile, one also requires that the HO potential is negligible at the boundary where the short-range interaction vanishes. 
In this sense, the external HO potential with weak oscillator strength $\omega$ is required in the applications of the MERE formula, where numerical tests are also necessary to verify the applicability of the MERE formula. There is no separate approximation based on the effective range of the potential.
As for the $np$ scattering, one of the authors has applied the MERE formula to solve the phase shifts in various channels utilizing realistic nuclear interactions \cite{Shirokov:2003kk,PhysRevC.79.014610} at CM scattering energies even beyond 50 MeV in a previous work \cite{PhysRevC.82.034003}.

We remark that our framework of phase shift calculations also works when the input eigenenergies have lower precision, which makes our framework applicable to near-term noisy quantum hardwares.
For this restricted model problem that serves to demonstrate the feasibility of the hybrid framework, we apply the limited data set in Table \ref{table:classical_and_numerical_energy_eigenvalues_of_np} for our phase shift calculations, where we retain input eigenenergies up to six significant figures. 
However, we found that the percentage differences of the phase shifts via our approach with respect to the analytical results are within 0.5$\%$ when retaining four significant figures for the input eigenenergies in the restricted data set [Table \ref{table:classical_and_numerical_energy_eigenvalues_of_np}]. 
Moreover, we also found fair agreement (the percentage difference being within a few percent) between our results and the analytical results when we retain only three significant figures for the input eigenenergies. We anticipate that this agreement may be improved by augmenting the input data set $\{ E_j(\omega) \}$ (achieved by including more eigenenergies of the scattering states and more choices of the trap strength values). Indeed, such an expanded data set would enable a robust error analysis of our approach. This will be addressed in future applications of our hybrid quantum-classical framework to the scatterings with complex nuclei. 

\subsection{Model problem 2: $n\alpha$ scattering}
\label{sec:model_problem_of_nalpha}

In this section, we discuss the phase shift of the elastic $n\alpha$ scattering in the $(1/2)^+$ channel.
In this problem, we take both nuclei to be point particles and adopt a phenomenological nuclear potential for the interaction between the $n$ and $\alpha$ particle in the problem. It is worth noting, however, that in the general applications, the colliding nuclei can have internal structures and the phenomenological nuclear potential is not available. In such complex case, we resort to the many-body formalism discussed in Sec. \ref{sec:many_nucleon_scattering}, where the inter-nucleus potential is obtained from the inter-nucleon forces that are derived from fundamental theories \cite{MACHLEIDT20111,PhysRevC.68.041001,Epelbaum:2008ga,Epelbaum:2014efa,Epelbaum:2014sza}. 

The Hamiltonian of the $n \alpha $ scattering system confined in the external HO potential of trap strength $\omega $ takes the form of Eq. \eqref{eq:relative_motion_of_XY} as
\begin{equation}
	H_{{\rm rel}, n\alpha }(\omega ) =  T_{{\rm rel},  n\alpha } + V_{{\rm int},  n\alpha } + V_{{\rm HO},  n\alpha }(\omega) ,
	\label{eq:confined_system_H_nAlpha}
\end{equation}
where we use the subscript ``$n \alpha $" to specify the system. We take the mass of the $\alpha $ particle to be $ 4 m= 3755.676 $ MeV and the reduced mass of the $n\alpha$ system to be $\mu _{n\alpha} = 4m/5 = 751.135$ MeV.\footnote{We neglect the binding energy of the $\alpha$ particle.} The interaction between the two nuclei is taken as the Woods-Saxon potential \cite{suhonen2007nucleons}
\begin{align}
	 V_{{\rm int},  n\alpha } = \frac{U_0}{1+e^{(r-R_0)/\alpha_0}}+(\vec{l} \cdot \vec{s})\frac{1}{r}\frac{\mathrm{d}}{\mathrm{d} r}\left[\frac{V_{ls}}{1+e^{(r-R_1)/\alpha_1}}\right]. \label{eq:Woods_Saxon_potential}
\end{align}
The parameters of the WS potential are $U_0= -43.0$ MeV, $R_0= 2.0$ fm, $\alpha_0=0.70$ fm, $V_{ls}=-40.0$ MeV$^2$, $R_1=1.5$ fm, and $\alpha_1=0.35$ fm \cite{BANG1979119,Shirokov:2016thl}. The term that is proportional to $ \vec{l} \cdot \vec{s} $ results from the contribution of the coupling between the orbital angular momentum $\vec{l}$ and spin $\vec{s}$; this term vanishes in our scattering problem that is in the uncoupled channel $(1/2)^+$. The HO potential $V_{{\rm HO},  n\alpha }(\omega) $ with weak trap strength presents finite corrections to the short-range inter-nucleus interaction $ V_{{\rm int},  n\alpha } $, where it also discretizes the continuum states. 

We solve the elastic scattering phase shift with our hybrid framework. To this end, we first solve the eigenenergies of the discretized scattering states of $H_{{\rm rel}, n\alpha }(\omega ) $ for a range of trap strengths via the quantum eigensolver. Based on the data set of eigenenergies and the trap strengths, we obtain the phase shift utilizing the MERE formula.

We follow the procedures presented in the $np$ problem for the practical calculations. We elect a discrete set of trap strengths $\omega \in \{ 1.72, \ 1.74, \ 1.76, \ 1.78, \ 1.80\}$ MeV. A sufficiently large HO basis set (with $N_{\rm max}=600$ at our elected value of $\Omega = 60$ MeV) is adopted to construct the matrix representation of $H_{{\rm rel}, n\alpha} (\omega )$ such that the eigenenergies of the low-lying discretized scattering states converge.

We solve the eigenenergies of $H_{{\rm rel}, n\alpha }(\omega ) $ via the rodeo algorithm. For simplicity, we demonstrate to solve the eigenenergies based on the effective Hamiltonian $H^{\rm eff}_{{\rm rel}, n\alpha }(\omega ) $, where 1) $H^{\rm eff}_{{\rm rel}, n\alpha }(\omega ) $ is constructed from the corresponding bare Hamiltonian $H_{{\rm rel}, n\alpha }(\omega ) $ via the Okubo-Lee-Suzuki method; and 2) $H^{\rm eff}_{{\rm rel}, n\alpha }(\omega ) $ retains the eigenenergies of an elected set of discretized scattering states of $H_{{\rm rel}, n\alpha }(\omega ) $. Although the choice of states is, in principle, arbitrary, we found it convenient to adopt the eigenenergies of the $1^{\rm st},\ 3^{\rm rd},\ 7^{\rm th}\ \text{and}\ 8^{\rm th}$ scattering states of the bare Hamiltonian $H_{{\rm rel}, n\alpha }(\omega ) $ for the purpose of our illustration. As such, the dimension of the effective Hamiltonian $H^{\rm eff}_{{\rm rel}, n\alpha }(\omega ) $ is 4 by 4.\footnote{We remark that it is understood that future application of the quantum eigensolver takes the full-size many-nucleon Hamiltonian as input. Such generalization is straightforward, while it is not the focus of this work.} The details of the implementation of the rodeo algorithm, which include the circuit construction, the choices of the random Gaussian variables, the preparation of the input state, the coarse and fine scans, and the fitting procedure for the peak centers, are the same as those in the $np$ example. 

The eigenenergies of the effective Hamiltonian $H^{\rm eff}_{{\rm rel}, n\alpha }(\omega ) $ via the rodeo algorithm are presented in Table \ref{table:classical_and_numerical_energy_eigenvalues_of_alphaN}. 
These are also the eigenenergies of the elected scattering states of the bare Hamiltonian $H_{{\rm rel}, n\alpha }(\omega ) $ according to our construction scheme of the effective Hamiltonians.
Our results via the rodeo algorithm agree well with the exact results that are obtained from convergence calculations via the straightforward matrix diagonalization of $H_{{\rm rel}, n\alpha }(\omega ) $ on classical computers. To achieve higher precision in the spectral calculations via the rodeo algorithm, we can improve our choices of the input states for larger overlap with the desired eigenstates [cf. Eq. \eqref{eq:inputState_implementation}], and increase the number of rodeo cycles. 

\renewcommand{\arraystretch}{1.5} 
\begin{table}[!ht]
	\setlength{\tabcolsep}{6pt}
	\centering  
		\caption{Eigenenergies $E_1$, $E_3$, $E_7$, and $E_8$ of the $1^{\rm st},\ 3^{\rm rd},\ 7^{\rm th}\ \text{and}\ 8^{\rm th}$ scattering states of the bare Hamiltonian $H_{{\rm rel}, n\alpha }(\omega ) $. The trap strength takes the value of $\omega = 1.72, \ 1.74, \ 1.76, \ 1.78,\ $and$\ 1.80 $ MeV. The results denoted as ``exact" are obtained by diagonalizing bare Hamiltonian $H_{{\rm rel}, n\alpha }(\omega ) $, while those obtained via the rodeo algorithm based on the corresponding effective Hamiltonians $H^{\rm eff}_{{\rm rel}, n\alpha }(\omega ) $ are denoted as ``rodeo". The eigenenergies are in the units of MeVs. See the text for more details.}  
		\label{table:classical_and_numerical_energy_eigenvalues_of_alphaN}  
		\begin{tabular}{ccccccccccc}  
			\hline
			\hline
			\multirow{2}{*}{Energy}&  
			\multicolumn{2}{c}{$\omega$ = 1.72 MeV}&\multicolumn{2}{c}{$\omega$ = 1.74 MeV}&\multicolumn{2}{c}{  $\omega$ = 1.76 MeV}&\multicolumn{2}{c}{$\omega$ = 1.78 MeV}&\multicolumn{2}{c}{$\omega$ = 1.80 MeV}\cr  
			\cline{2-3}\cline{4-5}\cline{6-7}\cline{8-9}\cline{10-11}
			& exact & rodeo & exact & rodeo & exact & rodeo & exact & rodeo & exact & rodeo \cr  
			\hline
			$E_1$&3.57377&3.57372 &3.62070&3.62077    &3.66771&3.66759    &3.71478&3.71482   &3.76194&3.76192\cr 
			$E_3$&10.9540&10.9538 &11.0869&11.0867    &11.2200&11.2199    &11.3531&11.3530    &11.4863&11.4860\cr 
			$E_7$&25.1461&25.1461 &25.4445&25.4445    &25.7429&25.7429    &26.0412&26.0412    &26.3395&26.3395\cr 
			$E_8$&28.6466&28.6466 &28.9851&28.9850    &29.3236&29.3238    &29.6622&29.6622    &30.0010&30.0010\cr 
			\hline 
			\hline
		\end{tabular}  
\end{table}  

\begin{figure}[ht]
	\centering
	\includegraphics[scale=0.4]{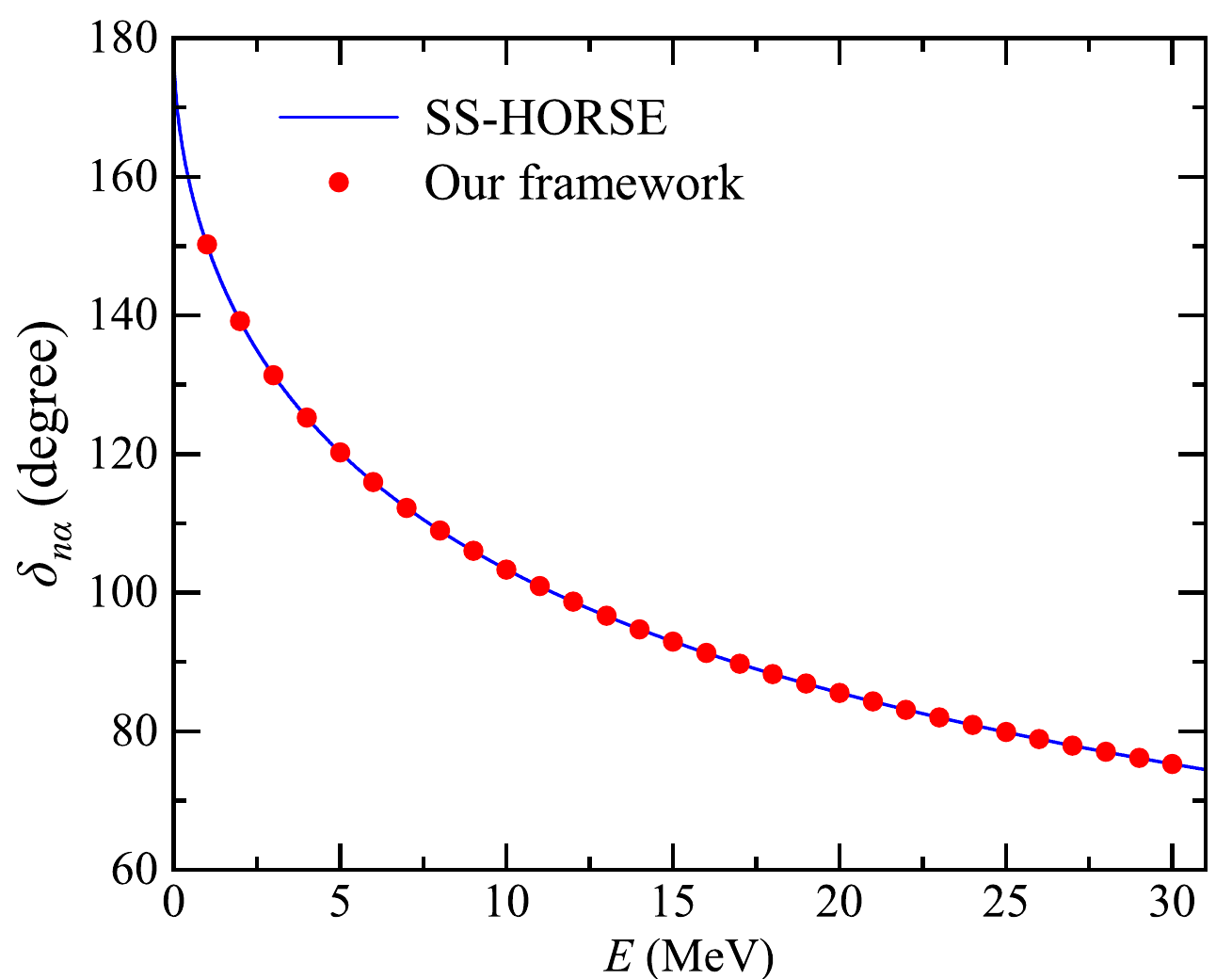}
	\caption{(color online) 
		The phase shift $\delta _{n\alpha } $ of the $n\alpha$ scattering in the $(1/2)^+$ channel as a function of the CM scattering energy $E$.
		The results obtained based on our framework is marked as red dots.
		For comparison, the results (blue solid line) computed by the SS-HORSE method \cite{Shirokov:2016thl} are presented as a function of $E$. 
	}
	\label{fig:final_results_alphaN}
\end{figure}

From the data set $\{ E_j ( \omega ) \}$ shown in Table \ref{table:classical_and_numerical_energy_eigenvalues_of_alphaN}, we utilize the MERE formula and extrapolate the phase shift $\delta _{n\alpha } $ of the elastic $n\alpha $ scattering in the $(1/2)^+$ channel on the classical computer following the procedures illustrated in Sec. \ref{sec:extrapolation_procedures_np}. These results are shown in Fig. \ref{fig:final_results_alphaN} as a function of the scattering energy in the CM frame. The results from our framework compare well with the calculations based on the SS-HORSE method [\citealp{Shirokov:2016thl}, Eq. (47) with parameters in Table II therein]: the difference between the results from our framework and those from the SS-HORSE method is below $1.7\%$ throughout the scattering energy $E \leq 30 $ MeV in the CM frame, while this difference is below $1\%$ for $E\leq 20$ MeV.\footnote{It is worth noting that we retain the scattering phase shift even beyond the scattering energy of $20.8$ MeV (which is the inelastic threshold of the $n\alpha$ scattering) in our realistic yet restricted model problem, where 1) we take the colliding nuclei to be point particles; and 2) we retain only one scattering channel of which the inter-nucleus potential is modeled by a simple phenomenological potential. With these simplifications, the inelastic scattering channels and other cross-channel scattering mechanisms are effectively excluded in the solution.} 
We tested that our hybrid framework provides useful phase shifts when we reduce the precision of the input eigenenergies (precision of phases shifts is comparable to the precision of the eigenenergies) even in the current demonstration problem with limited data set (Table \ref{table:classical_and_numerical_energy_eigenvalues_of_alphaN}). We expect that one may improve the precision and study the theoretical error of the results from our hybrid framework by retaining a more complete data set $\{ E_j ( \omega ) \}$ (i.e., more energy eigenvalues of the scattering states and more choices of $\omega$ values) for this model problem; the same is true for future applications of our framework to the full-scale many-body scattering problems that is sketched in Sec.  \ref{sec:sketch_of_the_framework}.

\section{Summary and outlook}
\label{sec:conclusion_and_outlook}

We present a hybrid quantum-classical framework to compute the elastic scattering phase shift of two well-bound nuclei in an uncoupled channel.
Within our framework, we introduce a many-nucleon formalism that enables the introduction of a harmonic oscillator (HO) potential of weak oscillator strength to modify the inter-nucleus interaction. 
We also propose the formalism to compute the eigenenergies of the low-lying discretized scattering states of the relative motion of the colliding nuclei as a function of the confining HO potential strength $\omega$ based on a set of many-nucleon structure calculations. 
With these eigenenergies of the relative motion as a function of $\omega $, we extract the elastic scattering phase shift in the uncoupled channel utilizing the formula of the modified effective range expansion (MERE) \cite{busch1998two, Stetcu_2007_PRA, Suzuki_2009, STETCU20101644, PhysRevC.82.034003,Blume_2012,PhysRevC.101.051602}. 

Our framework takes a hybrid approach.
The energy eigenvalues of the many-nucleon Hamiltonians are in general computationally challenging for classical computers, and we propose to calculate these eigenenergies via the quantum computing techniques that holds the promise to solve the many-nuclear structure problem with efficiency.
These spectral solutions are input to classical computers to obtain the continuum elastic scattering phase shift, where such post data processing are hard for quantum computers.  

We illustrate our framework via two limited model problems, where we take the colliding nuclei as point particles and model the inter-nucleus interactions by simple phenomenological potentials for explanatory purposes.
We show the application of the quantum eigensolver, which is elected to be the rodeo algorithm, to solve the eigenenergies of the low-lying discretized scattering states of the scattering systems confined in HO potentials.
The calculations via the rodeo algorithm are performed on the IBM Qiskit quantum simulator \cite{Qiskit}.
These results agree well with those from the classical calculations via straightforward matrix diagonalizations.
We also present the approach to extract the elastic scattering phase shifts. 
We find that the elastic scattering phase shifts obtained based on our framework agree well with the results from other theories.

Going forward, we plan to apply the framework to the elastic scatterings between two well-bound light nuclei in an uncoupled channel, e.g., the $n\alpha$ scattering in the $(3/2)^-$ channel below the inelastic threshold. 
The intrinsic motions of the colliding nuclei will be explicitly taken into account.
Hence we will proceed with the route $A1 \rightarrow A2 \rightarrow C1$ in Fig. \ref{fig:framework} to solve the scattering phase shift.
The difficulty is mainly the numerical solutions of the eigenenergies of the many-nucleon Hamiltonians. 
We plan to implement efficient quantum eigensolvers, such as the algorithm proposed in Refs. \cite{Du:2023bpw,Du:2024zvr,Liu:2024hmm}, for the desired spectral solutions of the relevant many-nucleon systems, which is to be cross-checked with those eigenenergies solved by the classical many-nucleon structure calculations, such as the no-core shell model \cite{Navratil:2000ww,Navratil:2000gs,Barrett:2013nh}.
These spectral solutions are the input to the MERE formula to extract the elastic scattering phase shift of interest.

\section{Acknowledgments}
We acknowledge fruitful discussions with Dean Lee, Andrey M. Shirokov, and Pieter Maris.
WD and JPV were partially supported by the U.S. Department of Energy under Grant DE-SC0023707(NuHaQ). P.W. and W.Z. were supported by the National Natural Science Foundation of China under Grant Nos. 12375117, 11975282, the Strategic Priority Research Program of Chinese Academy of Sciences under Grant No. XDB34000000.

\appendix

\section{Derivation of the MERE formula}
\label{sec:MERE_formula_derivation}

We provide the derivation of the MERE formula [Eq. \eqref{eq:MERE_formula} in the main text] utilizing the same techniques shown in Ref. \cite{Suzuki_2009}. 
While Ref. \cite{Suzuki_2009} treats the scattering between two particles of equal mass and does not show the MERE formula explicitly, we find it worthwhile to present the complete derivation for a more general setup with the scattering particles being of different masses. 
Interested readers are also referred to Ref. \cite{PhysRevC.101.051602} for a remedied version of the MERE formula.

We consider the scattering system of two point particles, which is confined in a weak external HO potential of strength $\omega $. In the single-particle coordinates, the Hamiltonian of the confined scattering system reads
\begin{equation}
	H _{2b}(\omega ) = \frac{\vec{p}_1^2}{2m_1} + \frac{1}{2} m_1 \omega \vec{r}_1^2 + \frac{\vec{p}_2^2}{2m_1} + \frac{1}{2} m_1 \omega \vec{r}_2^2 + V_{\rm int} . 
	\label{eq:B_totH12}
\end{equation}
The mass, momentum, and position of the $i^{\rm th}$ (with $i=1, \ 2$) particle are denoted by $m_i$, $\vec{p}_i$, and $\vec{r}_i$, respectively. $ V_{\rm int} $ denotes the inter-particle interaction. 

We rewrite $ H _{2b}(\omega ) $ in the CM frame. In particular, we take 
\begin{equation}
	\vec{R} = \frac{m_1 \vec{r}_1 + m_2 \vec{r}_2}{m_1 + m_2} , \ \vec{r} = \vec{r}_1 -\vec{r}_2 ,
\end{equation}
and Eq. \eqref{eq:B_totH12} becomes
\begin{eqnarray}
	H _{2b}(\omega )  = \underbrace{ \frac{ \vec{P}_{\rm CM} }{2M_{2b}} + \frac{1}{2} M_{2b} \omega ^2 \vec{R}^2 } _{\equiv H_{\rm CM} (\omega)} + \underbrace{ \frac{\vec{p}^2}{2\mu} + \frac{1}{2}\mu \omega ^2 \vec{r}^2 + V_{\rm int} }_{\equiv H (\omega)} ,
\end{eqnarray}
where we have $ M_{2b} = m_1 + m_2 $ and $\mu = (m_1m_2)/(m_1 + m_2)$. The momenta $\vec{P}_{\rm CM} $ and $ \vec{p} $ are conjugate to $ \vec{R} $ and $\vec{r}$, respectively. 

We factorize the CM motion and focus on the relative motion of the confined scattering system. The relative motion of the two particles is described by $ H (\omega) $ [Eq. \eqref{eq:confined_system}]. In the spherical coordinates, we have 
\begin{eqnarray}
	H (\omega) = -\frac{1}{2\mu r^2}\left[ 2r \frac{\partial}{\partial r} + r^2\frac{\partial^2}{\partial r^2} 
	- \hat{L}^2 \right]+\frac{1}{2}\mu\omega^2r^2  + V_{\rm int}, 
\end{eqnarray}
where the squared orbital angular momentum operator is
\begin{equation}
	\hat{L}^2 = - \frac{1	}{\sin \theta } \frac{\partial}{\partial \theta} \Big( \sin \theta \frac{\partial}{\partial \theta} \Big) - \frac{1	}{\sin ^2 \theta } \frac{\partial ^2}{\partial ^2 \varphi} .
\end{equation}

We consider the case where $ V_{\rm int} $ preserves the orbital angular momentum. 
With the separation of variables, we obtain the Schr\"odinger equation for the radial wave function $  \psi _l(r) = u_l(r) /r $ for the partial wave specified by the orbital angular momentum $l$, where $u_l(r)$ satisfies
\begin{equation}
	\frac{\partial^2}{\partial r^2}u_l(r)-\frac{l(l+1)}{r^2}u_l(r)- \mu^2\omega^2r^2 u_l(r)- 2\mu V_{\rm int} \ u_l(r) + 2\mu E u_l(r) = 0 ,
	\label{eq:schrodinger_equation_of_u_l_r}
\end{equation}
where $E $ denotes the eigenenergy of $H_{\rm rel } (\omega) $. Utilizing the dimensionless variable $x \equiv r/b_0 $ with $b_0 = \sqrt{1/(\mu \omega )}$, we rewrite Eq. \eqref{eq:schrodinger_equation_of_u_l_r} as
\begin{equation}
	\frac{\partial^2}{\partial x^2}u_l(x)-\frac{l(l+1)}{x^2}u_l(x)-x^2u_l(x)-\frac{2V_{\rm int}}{\omega}u_l(x)+\eta u_l(x)=0, 
	\label{eq:schrodinger_equation_of_u_l_x}
\end{equation}
with $\eta =\frac{2E }{\omega}$. 

To proceed, we {\it assume} that (1) $ V_{\rm int} $ is a short-range potential; we take $V_{\rm int} \neq 0 $ for $ r < s_0 $ and $V_{\rm int}= 0 $ for $r\geq s_0 $, with $s_0$ being small; (2) the potential $V_{\rm HO} = \frac{1}{2}\mu \omega ^2 \vec{r}^2 \ll V_{\rm int} $ for $r < s_0 $; and (3) the oscillator potential is small and can still be neglected in the limit $r\rightarrow s_0^+$. Note that these {\it assumptions} need to be numerically examined for specific implementations with different interactions $V_{\rm int}$ and partial waves \cite{Suzuki_2009,PhysRevC.82.034003}. 

With these assumptions, we now try to prove the MERE formula. In the limit $r\rightarrow s_0^+$, the oscillator potential is still weak such that one can neglect it. We have the scattering solution of Eq. \eqref{eq:schrodinger_equation_of_u_l_r} with the positive energy $E = {p^2}/{(2\mu)}$ to be
\begin{equation}
	u_l(r) = \mathcal{B} _l pr[j_l(pr)-n_l(pr)\tan\delta_l], 
	\label{eq:wave function_r_outside_of_interaction}
\end{equation}
where $\delta_l $ is the phase shift. $\mathcal{B} _l$ is the normalization factor. $j_l(pr) $ and $n_l(pr)$ are the spherical Bessel and Neumann functions \cite{arfken2013mathematical}, respectively. With the assumptions mentioned above, we have $ r \approx 0$ in the vicinity of $s_0^+$. Therefore, we have 
\begin{equation}
	u_l(r) \rightarrow \mathcal{B} ' _l r^{l+1} \Big[ 1+(2l+1)[(2l-1)!!]^2\frac{\tan \delta_l}{(pr)^{2l+1}} \Big] ,
	\label{eq:boarder_1}
\end{equation}
when $pr \rightarrow 0$. The coefficient $\mathcal{B} ' _l$ can be expressed as $ \mathcal{B} ' _l = \mathcal{B}  _l p^l / [(2l+1)!!] $. In achieving Eq. \eqref{eq:boarder_1}, we have applied the identities 
\begin{equation}
	j_l(x) \rightarrow x^l /[(2l+1)!!] , \  n_l(x) \rightarrow -\frac{(2l+1)!!}{x^{l+1}} ,
\end{equation}
as $x\rightarrow 0$. 

For $r > s_0$, we have $V_{\rm int} = 0$ and Eq. \eqref{eq:schrodinger_equation_of_u_l_x} reduces to 
\begin{equation}
	\frac{\partial^2}{\partial x^2}u_l(x)-\frac{l(l+1)}{x^2}u_l(x)-x^2u_l(x)+\eta u_l(x)=0.  \label{eq:schrodinger_equation_of_u_l_x_without_interacting}
\end{equation}
The solution of Eq. \eqref{eq:schrodinger_equation_of_u_l_x_without_interacting} admits the form \cite{Suzuki_2009}
\begin{equation}
	u_l(x)=e^{-x^2/2}\left[  c_1'x^{l+1}M\left( \frac{2l+3-\eta }{4}, l+\frac{3}{2};x^2  \right)  +c_2'x^{-l}M\left( \frac{1-2l-\eta }{4}, \frac{1}{2}-l;x^2  \right)   \right] ,
	\label{eq:wave_function_without_V}
\end{equation}
where $M(a, c; z)$ denotes the confluent hypergeometric function \cite{arfken2013mathematical}. $c'_1$ and $c'_2$ are the coefficients.
As $ z \rightarrow \infty $, we have \cite{Mathews_2022}
\begin{equation}
	M(a,c;z) \rightarrow \frac{\Gamma(c)}{\Gamma(a)}z^{a-c}e^z 
	\label{CHF_when_z_is_very_large} .
\end{equation}
Therefore, as $x\rightarrow \infty$, we have 
\begin{equation}
	u_l (x) \rightarrow \Big[ c_1'\frac{\Gamma(3/2+l)}{\Gamma((3+2l-\eta )/4)} +c_2'\frac{\Gamma(1/2-l)}{\Gamma((1-2l-\eta )/4)} \Big]x^{-(1+ \eta )}e^{x^2/2}. 
	\label{wave_function_when_z_is_very_large}
\end{equation}
Since we require $u_l(x) \rightarrow 0$ for $x\rightarrow \infty $, the ratio of $c'_1$ and $c'_2$ should satisfy 
\begin{equation}
	\frac{c_2'}{c_1'}=-\frac{\Gamma(3/2+l)\Gamma((1-2l-\eta )/4)}{\Gamma(1/2-l)\Gamma((3+2l-\eta )/4)}. 
	\label{eq:ratio_1}
\end{equation}

With the identities 
\begin{equation}
	\Gamma \Big(\frac{1}{2}+ l \Big) = \sqrt{\pi}\frac{(2l-1)!!}{2^l}, \ \Gamma\Big(\frac{1}{2}-l\Big) =  \sqrt{\pi}\frac{(-2)^l}{(2l-1)!!},
	\label{properties_of_gamma_function}
\end{equation}
we can simplify Eq. \eqref{eq:ratio_1} as
\begin{equation}
	\frac{c_2'}{c_1'}=(-1)^{l+1} \Big(l+\frac{1}{2} \Big)
	\frac{[(2l-1)!!]^2}{2^{2l}}\frac{\Gamma((1-2l- \eta )/4)}{\Gamma((3+2l-\eta )/4)}. 
	\label{the_ratio_of_coefficients_1}
\end{equation} 

With $x \rightarrow s_0^+$, we have from Eq. \eqref{eq:wave_function_without_V} that
\begin{equation}
	u_l(x) \rightarrow c_1'x^{l+1}\left[ 1+\frac{c_2'}{c_1'}x^{-(2l+1)} \right] , 
	\label{eq:boarder_2}
\end{equation}
where we have applied the identity $M(a, c; z) \rightarrow 1$ as $z \rightarrow 0$.

We match the solutions of the radial wave function at $r \rightarrow s^+_0$. By comparing Eqs. \eqref{eq:boarder_1} and \eqref{eq:boarder_2}, we obtain
\begin{equation}
	\frac{c_2'}{c_1'}=(2l+1)[(2l-1)!!]^2\left( \sqrt{{\mu\omega}} \right)^{2l+1}\frac{\tan\delta_l}{p^{2l+1}}. 
	\label{eq:ratio_2}
\end{equation}

Finally, the ratio ${c_2'}/{c_1'}$ should satisfy both Eq. \eqref{eq:ratio_1} and Eq. \eqref{eq:ratio_2}. By equating  Eqs. \eqref{eq:ratio_1} and \eqref{eq:ratio_2}, we obtain 
\begin{equation}
	p^{2l+1}\cot\delta_l=(-1)^{l+1}\left(4\mu\omega\right)^{l+\frac{1}{2}}\frac{\Gamma\left( \frac{3+2l}{4}-\frac{E }{2\omega} \right)}{\Gamma\left( \frac{1-2l}{4}-\frac{E }{2\omega} \right)} ,
	\label{eq:final_result_2}
\end{equation}
where we recall $p = \sqrt{2\mu E }$ and $\eta  = 2E /\omega $. This completes the derivation of Eq. \eqref{eq:MERE_formula} in the main text.

\section{Hamiltonian of the scattering system in harmonic oscillator potential}
\label{sec:appendix_MBHamiltonian}

\begin{figure}[!ht]
	\centering
	\includegraphics[scale=0.45]{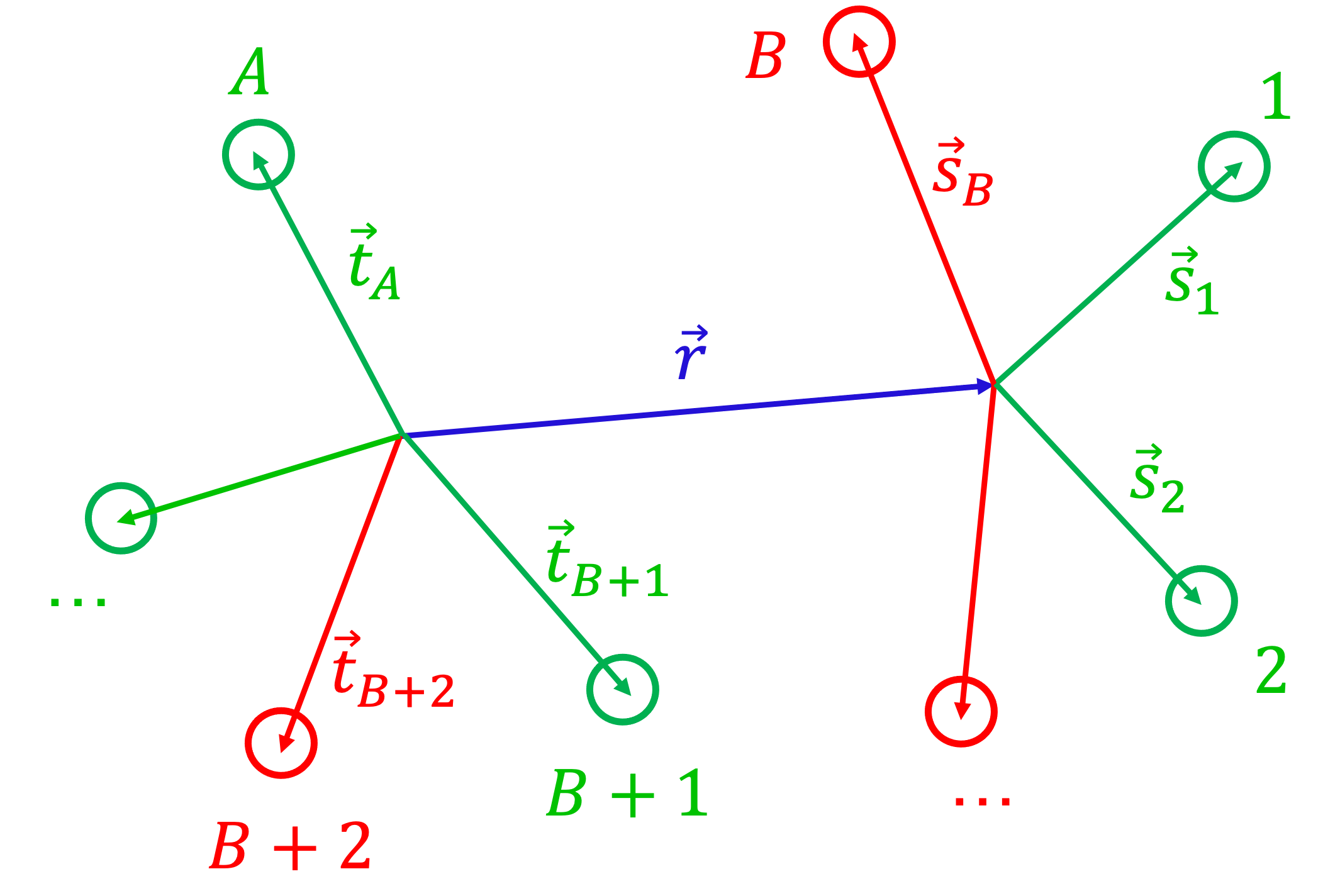}
	\caption{(color online) The two-cluster configuration of the $A$-nucleon system. The right cluster consists of $B$ nucleons and the left cluster contains $(A-B)$ nucleons. See the text for more details.
	}
	\label{fig:clusterConfigration}
\end{figure}

According to the discussion in the main text, the $A$-nucleon Hamiltonian with the additional harmonic oscillator (HO) potential acting on pairwise nucleons is [Eq. \eqref{eq:many_nucleon_Hamiltonian_in_trap}]
\begin{align}
	H_A(\omega ) = \frac{1}{2Am} \sum _{i<j}^A (\vec{p}_i - \vec{p}_j)^2 + \sum _{i<j}^A V_{ij} + \frac{1}{2A} m \omega ^2 \sum _{i<j}^A (\vec{r}_i - \vec{r}_j)^2 ,
	\label{eq:appendixEq_A}
\end{align}
where $\vec{r}_i$ and $\vec{p}_i$ denotes the single-nucleon position and momentum, respectively. We can derive the Hamiltonian of a two-cluster configuration of the $A$-nucleon system from $H_A(\omega )$. 

As shown in Fig. \ref{fig:clusterConfigration}, we define $\vec{s}_i$ with $i \in [1,B]$ to be the position of the $i^{\rm th}$ nucleon with respect to the mass center $\vec{R}_{\rm right} = \frac{1}{B} \sum _{i=1}^B \vec{r}_i$ of the $B$-nucleon cluster (we assume the neutron and proton are of equal mass). Meanwhile, $\vec{t}_j$ with $j \in [B+1,A]$ is the position of the $j^{\rm th}$ nucleon with respect to the mass center $\vec{R}_{\rm left} = \frac{1}{A-B} \sum _{j=B+1}^A \vec{r}_j$ of the $(A-B)$-nucleon cluster. The relative position vector $\vec{r}$ from the mass center of the left cluster to that of the right cluster is defined as $\vec{r} = \vec{R}_{\rm right} - \vec{R}_{\rm left}$. Therefore, we have 
\begin{align}
	\begin{cases}
		\ \vec{r}_i - \vec{r}_j =  \vec{s}_i - \vec{s}_j, \ \ & \text{for} \ \ \ i \in [1,B], \ \  \text{and} \ \  j \in [1,B] , \\
		\ \vec{r}_i - \vec{r}_j = -\vec{t}_j + \vec{r} + \vec{s}_i , \ \ & \text{for} \ \ \ i \in [1, B], \ \  \text{and} \ \  j \in [B+1,A] , \\
		\ \vec{r}_i - \vec{r}_j =  \vec{t}_i - \vec{t}_j, \ \ & \text{for} \ \ \ i \in [B+1,A], \ \  \text{and} \ \  j \in [B+1,A] .
	\end{cases}
\end{align}
Hence, we have the following identity
\begin{align}
	\sum _{i<j}^A (\vec{r}_i - \vec{r}_j)^2 = & \sum _{i=1, i<j}^B (\vec{r}_i - \vec{r}_j)^2 + \sum _{i=B+1, i<j} ^A (\vec{r}_i - \vec{r}_j)^2 + \sum _{i=1}^B \sum _{j=B+1}^A (\vec{r}_i - \vec{r}_j)^2 \nonumber \\
	= & \sum _{i=1, i<j}^B (\vec{s}_i - \vec{s}_j)^2 + \sum _{i=B+1, i<j} ^A (\vec{t}_i - \vec{t}_j)^2 + (A-B) \sum _{i=1}^B \vec{s}_i^2 + B(A-B) \vec{r}^2 + B \sum _{i=B+1}^A \vec{t}^2_i .
\end{align}

Similarly, we define 
\begin{align}
	\begin{cases}
		\ \vec{p}_i - \vec{p}_j =  m  \dot{\vec{s} } _i - m \dot{\vec{s} } _j   , \ \ & \text{for} \ \ \ i \in [1,B], \ \  \text{and} \ \  j \in [1,B] , \\
		\ \vec{p}_i - \vec{p}_j =  m  \dot{\vec{s} } _i + m \dot{\vec{r} } - m \dot{\vec{t} } _j , \ \ & \text{for} \ \ \ i \in [1,B], \ \  \text{and} \ \  j \in [B+1,A] , \\
		\ \vec{p}_i - \vec{p}_j =  m  \dot{\vec{t} } _i - m \dot{\vec{t} } _j , \ \ & \text{for} \ \ \ i \in [B+1,A], \ \  \text{and} \ \  j \in [B+1,A] .
	\end{cases}
\end{align}
Then, we have
\begin{align}
	\sum _{i<j}^A (\vec{p}_i - \vec{p}_j)^2 = & \sum _{i=1, i<j}^B (\vec{p}_i - \vec{p}_j)^2 + \sum _{i=B+1, i<j} ^A (\vec{p}_i - \vec{p}_j)^2 + \sum _{i=1}^B \sum _{j=B+1}^A (\vec{p}_i - \vec{p}_j)^2 \\
	= & \sum _{i=1, i<j}^B (m  \dot{\vec{s} } _i - m \dot{\vec{s} } _j )^2 + \sum _{i=B+1, i<j} ^A (  m  \dot{\vec{t} } _i - m \dot{\vec{t} } _j )^2 + m^2 (A-B) \sum _{i=1}^B \dot{\vec{s} } _i^2  +  B(A-B) m^2 \dot{\vec{r} }^2 + B m^2 \sum _{i=B+1}^A \dot{\vec{t} } _i^2 .
\end{align}

The inter-nucleon interaction terms can also be sorted according to the clusters
\begin{align}
	\sum _{i<j}^A V_{ij} = \sum _{i=1, i<j}^B V_{ij} + \sum _{i=B+1, i<j} ^A  V_{ij} + \sum _{i=1}^B \sum _{j=B+1}^A V_{ij} .
\end{align}

With the above definitions, we can rewrite Eq. \eqref{eq:appendixEq_A} as
\begin{align}
	& H_A(\omega ) \nonumber \\
	= &  \frac{1}{2Am}  \left\{  \sum _{i=1, i<j}^B (m  \dot{\vec{s} } _i - m \dot{\vec{s} } _j )^2 + \sum _{i=B+1, i<j} ^A (  m  \dot{\vec{t} } _i - m \dot{\vec{t} } _j )^2 + m^2 (A-B) \sum _{i=1}^B \dot{\vec{s} } _i^2  +  B(A-B) m^2 \dot{\vec{r} }^2 + B m^2 \sum _{i=B+1}^A \dot{\vec{t} } _i^2 \right\} \nonumber \\
	& +  \sum _{i=1, i<j}^B V_{ij} + \sum _{i=B+1, i<j} ^A  V_{ij} + \sum _{i=1}^B \sum _{j=B+1}^A V_{ij} \nonumber \\
	&  + \frac{1}{2A} m \omega ^2  \left\{   \sum _{i=1, i<j}^B (\vec{s}_i - \vec{s}_j)^2 + \sum _{i=B+1, i<j} ^A (\vec{t}_i - \vec{t}_j)^2 + (A-B) \sum _{i=1}^B \vec{s}_i^2 + B(A-B) \vec{r}^2 + B \sum _{i=B+1}^A \vec{t}^2_i \right\} .
\end{align}
By sorting the terms according to the nucleons in left cluster and/or right cluster, we have
\begin{align}
	& H_A(\omega ) \nonumber \\
	=& \frac{1}{2Am}  B(A-B) m^2 \dot{\vec{r} }^2 + \frac{1}{2A} m \omega ^2  B(A-B) \vec{r}^2 + \sum _{i=1}^B \sum _{j=B+1}^A V_{ij} \nonumber \\
	& + \frac{1}{2Am} \sum _{i=1, i<j}^B (m  \dot{\vec{s} } _i - m \dot{\vec{s} } _j )^2 + \frac{1}{2A} m \omega ^2 \sum _{i=1, i<j}^B (\vec{s}_i - \vec{s}_j)^2 +  \sum _{i=1, i<j}^B V_{ij} + \frac{A-B}{2Am}  m^2  \sum _{i=1}^B \dot{\vec{s} } _i^2 + \frac{A-B}{2A} m \omega ^2  \sum _{i=1}^B \vec{s}_i^2 \nonumber \\
	& + \frac{1}{2Am} \sum _{i=B+1, i<j} ^A (  m  \dot{\vec{t} } _i - m \dot{\vec{t} } _j )^2 + \frac{m \omega ^2}{2A}   \sum _{i=B+1, i<j} ^A (\vec{t}_i - \vec{t}_j)^2 + \sum _{i=B+1, i<j} ^A  V_{ij} + \frac{B m^2}{2Am} \sum _{i=B+1}^A \dot{\vec{t} } _i^2 + \frac{ B m \omega ^2}{2A}  \sum _{i=B+1}^A \vec{t}^2_i .
	\label{eq:app_A10}
\end{align}
We define the reduced mass of the two clusters as $ \mu = B(A-B) m/A$ and the intrinsic momentum between the clusters to be $\vec{p} = \mu \dot{\vec{r}}$. Meanwhile, we define $\vec{k} _i = m \dot{\vec{s}}_i$ with $i \in [1, B]$ to be the single-nucleon momentum in the CM frame of the right cluster, and $\vec{q} _i = m \dot{\vec{t}}_i$ with $i \in [B+1, A]$ to be the single-nucleon momentum in the CM frame of the left cluster. The above equation becomes
\begin{align}
	H_A(\omega ) = & 
	\frac{1}{2\mu } \vec{p}^2 + \frac{1}{2} \mu \omega ^2 \vec{r}^2 + \sum _{i=1}^B \sum _{j=B+1}^A V_{ij}  \nonumber \\
	&  + \underbrace{ \frac{1}{2Am}  \sum _{i=1, i<j}^B (\vec{k}_i - \vec{k}_j)^2 + \frac{1}{2A} m \omega ^2 \sum _{i=1, i<j}^B (\vec{s}_i - \vec{s}_j)^2 +  \sum _{i=1, i<j}^B V_{ij} + \frac{A-B}{2Am} \sum _{i=1}^B \vec{k}_i^2 + \frac{A-B}{2A} m \omega ^2 \sum _{i=1}^B \vec{s}_i^2 }_{ \equiv H_{ \mathcal{X}'} (\omega)} \nonumber \\
	&  + \underbrace{ \frac{1}{2Am} \sum _{i=B+1, i<j} ^A (\vec{q}_i - \vec{q}_j)^2 + \frac{1}{2A} m \omega ^2 \sum _{i=B+1, i<j} ^A (\vec{t}_i - \vec{t}_j)^2 + \sum _{i=B+1, i<j} ^A  V_{ij} + \frac{B}{2Am} \sum _{i=B+1}^A \vec{q}^2_i + \frac{B}{2A} m \omega ^2 \sum _{i=B+1}^A \vec{t}^2_i }_ {\equiv H_{ \mathcal{Y}'} (\omega)} . 
	\label{eq:appendixEq10}
\end{align}
We can write the above equation as
\begin{align}
	H_A(\omega ) = H_{\rm rel} (\omega ) + H_{ \mathcal{X}'} (\omega) + H_{ \mathcal{Y}' } (\omega).
	\label{eq:clusterization_2}
\end{align}
$H_{\rm rel} (\omega ) $ denotes the intrinsic motion of the two clusters
\begin{equation}
	H_{\rm rel} (\omega ) =  \frac{1}{2\mu } \vec{p}^2 + \frac{1}{2} \mu \omega ^2 \vec{r}^2 + \sum _{i=1}^B \sum _{j=B+1}^A V_{ij} ,
	\label{eq:Ham_HAB_rel}
\end{equation}
where $ \sum _{i=1}^B \sum _{j=B+1}^A V_{ij} $ denotes the inter-cluster interaction, which can be modelled by phenomenological approaches (e.g., the Woods-Saxon potential in the example of the $n\alpha$ scattering in Sec. \ref{sec:model_problem_of_nalpha}). The second and third lines of Eq. \eqref{eq:appendixEq10} can be understood as the intrinsic Hamiltonians $H_{ \mathcal{X}'} (\omega)$ and $ H_{ \mathcal{Y}'} (\omega) $ of the confined $B$- and $(A-B)$-nucleon clusters in respective nuclear environments represented as scaled harmonic oscillators (up to proper rescaling of the single nucleon mass $m$ and the trap strength $\omega $). 

Furthermore, we can rewrite $H_{ \mathcal{X}'} (\omega)$ and $ H_{ \mathcal{Y}'} (\omega) $ with proper rescalings of $m$ and $\omega$. In particular, by defining $m' = Am/B $ and $\omega ' = B\omega /A$, $H_{ \mathcal{X}'} (\omega)$ can be written as
\begin{multline}
	 H_{ \mathcal{X}'} (\omega) = 
	\frac{1}{2Bm'}  \sum _{i=1, i<j}^B (\vec{k}_i - \vec{k}_j)^2 + \frac{m' (\omega ') ^2}{2B}  \sum _{i=1, i<j}^B (\vec{s}_i - \vec{s}_j)^2 +  \sum _{i=1, i<j}^B V_{ij} \\
	+ \frac{A-B}{B} \left[ \frac{1}{2m'} \sum _{i=1}^B \vec{k}_i^2 + \frac{1}{2} m' (\omega ') ^2 \sum _{i=1}^B \vec{s}_i^2 \right] .
	\label{eq:Ham_HB}
\end{multline}

Similarly, if we define  $m'' = Am/(A-B) $ and $\omega '' = (A-B)\omega /A$, the cluster Hamiltonian $ H_{ \mathcal{Y}'} (\omega) $ can be written as
\begin{multline}
	 H_{ \mathcal{Y}'} (\omega ) =
	\frac{1}{2(A-B)m''} \sum _{i=B+1, i<j} ^A (\vec{q}_i - \vec{q}_j)^2 + \frac{1}{2(A-B)} m'' (\omega'') ^2 \sum _{i=B+1, i<j} ^A (\vec{t}_i - \vec{t}_j)^2 + \sum _{i=B+1, i<j} ^A  V_{ij} \\
	+ \frac{B}{A-B} \left[ \frac{1}{2m''} \sum _{i=B+1}^A \vec{q}^2_i 
	+  \frac{1}{2} m'' (\omega '') ^2 \sum _{i=B+1}^A \vec{t}^2_i \right] .
\label{eq:Ham_HA_B}
\end{multline}

We note that the first three terms in $H_{ \mathcal{X}'} (\omega ) $ represent the form of Eq. \eqref{eq:appendixEq_A} for the $B$-nucleon system. The last term in $H_{ \mathcal{X}'} (\omega ) $ denotes the effect of the nuclear environment, which takes the form of the pure HO with the scaled mass $m'$ and trap strength $\omega '$. Analogous statements hold for $H_{ \mathcal{Y}'} (\omega ) $.

\subsection{Special cases}
As a special case with $B = A-1$, the reduced mass is $ \mu  = (A-1) m/A $. Eq. \eqref{eq:appendixEq10} becomes
\begin{align}
	H_A(\omega ) = & \frac{1}{2 \mu} \vec{p}^2 + \frac{1}{2}  \mu  \omega ^2 \vec{r}^2 + \sum _{i=1}^{A-1}  V_{iA} \nonumber \\
	& + \frac{1}{2Am}  \sum _{i=1, i<j}^{A-1} (\vec{k}_i - \vec{k}_j)^2 + \frac{1}{2A} m \omega ^2 \sum _{i=1, i<j}^{A-1} (\vec{s}_i - \vec{s}_j)^2 +  \sum _{i=1, i<j}^{A-1} V_{ij} + \frac{1}{2Am} \sum _{i=1}^{A-1} \vec{k}_i^2 + \frac{1}{2A} m \omega ^2 \sum _{i=1}^{A-1} \vec{s}_i^2 .
\end{align}
We remark that the terms corresponding to the intrinsic motion of the one-nucleon cluster vanish in this cases (note that there are no intrinsic coordinates for the one-nucleon cluster). With the scaled nucleon mass $m''' = Am/(A-1)$ and the scaled harmonic oscillator potential strength $\omega ''' = (A-1)\omega /A$, the second line of the above equation can be written as
\begin{align}
	H_{\mathcal{X}'}(\omega ) =  &  \frac{1}{2(A-1)m'''}  \sum _{i=1, i<j}^{A-1} (\vec{k}_i - \vec{k}_j)^2 + \frac{1}{2(A-1)} m''' (\omega ''') ^2 \sum _{i=1, i<j}^{A-1} (\vec{s}_i - \vec{s}_j)^2 +  \sum _{i=1, i<j}^{A-1} V_{ij} \nonumber \\
	& +  \frac{1}{A-1} \left\{ \frac{1}{2m'''} \sum _{i=1}^{A-1} \vec{k}_i^2 + \frac{1}{2} m''' (\omega''') ^2 \sum _{i=1}^{A-1} \vec{s}_i^2 \right\}  .
\end{align}
We note that the above equation presents the Hamiltonian of the $(A-1)$-nucleon system (with the mass being $m'''$) in the presence of the HO potential (with the strength being $\omega '''$) acting between each pair of nucleons, where the effect of the nuclear environment to each nucleon can be considered as a scaled HO. 

Moreover, in the case of the neutron-neutron and neutron-proton scatterings, the above formalism sees further simplifications. By direct observations, we only expect the terms of $H_{\rm rel} (\omega)$ in Eq. \eqref{eq:clusterization_2} (the two clusters are both one-nucleon systems and neither cluster has intrinsic motion).

\section{Constraint terms for many-nucleon Hamiltonians}
\label{sec:LL_terms_MN_Ham}

We discuss that we can extract the elastic scattering phase shift between two well-bound nuclei $\mathcal{X}$ and $\mathcal{Y}$ utilizing the MERE formula [Eq. \eqref{eq:MERE_formula}] based on the low-lying eigenenergies of $H_{\rm rel} (\omega ) $, which are inferred from those of $ H_A(\omega ) $, $H_{ \mathcal{X}} (\omega') $ [Eq. \eqref{eq:Ham_HB}], and $H_{ \mathcal{Y}'} (\omega ) $ [Eq. \eqref{eq:Ham_HA_B}] according to Eq. \eqref{eq:energy_decomposition_scheme}. These eigenenergies can be computed via {\it ab initio} many-nucleon structure calculations. In this section, we discuss two techniques that facilitate the structure calculations of the $\mathcal{N}$-nucleon problem with $\mathcal{N} \in \{A,\ B,\ A-B \}$ and the corresponding Hamiltonian $ H_{\mathcal{N}} \in \{ H_A,\ H_{ \mathcal{X}'}, \  H_{ \mathcal{Y}'}   \}$.

The first technique aims to isolate the spurious CM excited states from the states without CM excitation in many-nucleon calculations utilizing the single-nucleon bases. Such spurious CM excitations result from our adoption of the Slater determinant bases and from the fact that the Hamiltonian of the CM motion of the $\mathcal{N}$-nucleon system commutes with the intrinsic Hamiltonian $H_{\mathcal{N}}(\omega) $ of the system. The spectral solutions without CM excitation can be isolated by introducing an additional Lawson-Lipkin term \cite{Lipkin:1958,Gloeckner:1974sst}
\begin{equation}
	W_{\mathcal{N}}^{\rm CM} (\Lambda _{\rm CM} , \Omega) = \Lambda _{\rm CM} (H^{\rm CM}_{\mathcal{N}} - \frac{3}{2} \Omega )
	\label{eq:ll_term_CM}
\end{equation}
to $H_{\mathcal{N}} $. The CM Hamiltonian of the $\mathcal{N}$-nucleon system is \cite{Barrett:2013nh}
\begin{equation}
	H^{\rm CM}_{\mathcal{N}} = T^{\rm CM}_{\mathcal{N}} + U^{\rm CM}_{\mathcal{N}},
\end{equation}
with 
\begin{align}
	T^{\rm CM}_{\mathcal{N}} =& \frac{1}{2\mathcal{N} m} \Big(\sum ^{\mathcal{N}}_i \vec{p}_i \Big)^2 = \frac{1}{2\mathcal{N} m} \sum ^{\mathcal{N}}_{i,j} \vec{p}_i \cdot \vec{p}_j , \\
	U^{\rm CM}_{\mathcal{N}} =& \frac{1}{2} \mathcal{N} m \Omega ^2 \Big[ \frac{1}{\mathcal{N}} \sum ^{\mathcal{N}}_i \vec{r}_i \Big]^2 = \frac{1}{2\mathcal{N} }m \Omega ^2 \sum ^{\mathcal{N}}_{i,j} \vec{r}_i \cdot \vec{r}_j ,
\end{align}
where $\Omega $ denotes the oscillator strength of the HO basis. With sufficiently large and positive coefficient $\Lambda _{\rm CM} $, the states with CM excitations are shifted up and away from the low-lying spectrum.

The second technique aims to isolate the many-nucleon states of different total angular momenta $J _{\mathcal{N}}$ for investigating specific scattering channels. The operators $\hat{J}^2 $, $\hat{M}_{J}$, $H_{\mathcal{N}} $, and $H^{\rm CM}_{\mathcal{N}} $ are mutually commuting. For the energy eigenstate $ | \Psi _{\mathcal{N}} \rangle $ of $H_{\mathcal{N}} $ and $H^{\rm CM}_{\mathcal{N}} $, we have 
\begin{equation}
	\hat{J}^2 | \Psi _{\mathcal{N}} \rangle = J_{\mathcal{N}} (J_{\mathcal{N}} +1) | \Psi  _{\mathcal{N}} \rangle, \ \hat{M}_{J} | \Psi  _{\mathcal{N}} \rangle = M_{J_{\mathcal{N}}} | \Psi  _{\mathcal{N}} \rangle ,
\end{equation}
where $ M_{J_{\mathcal{N}}} $ denotes the projection of $ J_{\mathcal{N}} $. Note also that the energy eigenstates in the many-nucleon structure calculations are degenerate in $ M_{J_{\mathcal{N}}} $. It is possible to retain only those Slater determinant bases with $ M_{J_{\mathcal{N}}} = J_{\mathcal{N}} $ in constructing the basis space for the many-nucleon structure calculation. With this basis construction scheme, we can eliminate all the possible energy eigenstates with $J_{\mathcal{N}} < M_{J_{\mathcal{N}}} $ in the spectral solution. We then introduce another constraint term
\begin{align}
	W^J_{\mathcal{N}} (\Lambda _J, J_{\mathcal{N}}) = \Lambda _J [ \hat{J}^2 - J_{\mathcal{N}}(J_{\mathcal{N}} + 1) ] ,
\end{align}
to the many-nucleon Hamiltonian $H_{\mathcal{N}} $. With $\Lambda _J $ being a sufficiently large positive number, the action of $ W^J_{\mathcal{N}} (\Lambda _J, J_{\mathcal{N}}) $ lifts the scattering states of which the total angular momentum is not $J_{\mathcal{N}}$.
This facilitates isolating the scattering states according to the $J_{\mathcal{N}}$ value of the desired scattering channel in the low-lying spectrum.

\normalem
\bibliography{apssamp}
\end{document}